\documentclass[fleqn,usenatbib]{mnras}

\usepackage{newtxtext,newtxmath}
\usepackage[T1]{fontenc}

\DeclareRobustCommand{\VAN}[3]{#2}
\let\VANthebibliography\thebibliography
\def\thebibliography{\DeclareRobustCommand{\VAN}[3]{##3}\VANthebibliography}

\usepackage{graphicx}	% Including figure files
\usepackage{amsmath}	% Advanced maths commands
\usepackage{mathtools}
\usepackage{booktabs}
\usepackage{longtable}
\usepackage{cuted}
\usepackage{placeins}
\usepackage{enumitem}
\setlist[enumerate]{itemsep=0mm}
\usepackage{float}
\usepackage{soul}

\title[Pair Production in GRMHD Simulations]{Electron-positron Pair Production in Global GRMHD Simulations of Black Hole Accretion Flows}

\author[H. S. Chan et al.]{
Ho-Sang Chan,$^{1,2}$\thanks{E-mail: hschanastrophy1997@gmail.com}
Jason Dexter,$^{1,2}$
Mitchell C. Begelman$^{1,2}$
\\
$^{1}$JILA, University of Colorado and National Institute of Standards and Technology, 440 UCB, Boulder, CO 80309-0440, USA\\
$^{2}$Department of Astrophysical and Planetary Sciences, University of Colorado, 391 UCB, Boulder, CO 80309, USA
}

\date{Accepted 2026 February 26. Received 2026 January 31; in original form 2025 November 7}

\pubyear{\the\year{}}

\begin{document}
\label{firstpage}
\pagerange{\pageref{firstpage}--\pageref{lastpage}}
\maketitle

% Abstract of the paper
%W
\begin{abstract}
We present global, three-dimensional general relativistic magnetohydrodynamic simulations of accreting black holes that incorporate pair physics. Pairs are modeled as a passive scalar that maintains a constant temperature. For high accretion rate models, we observe a maximum pair fraction of $\sim \mathcal{O}(0.01)$, consistent with those inferred from some X-ray binaries, and identify a `pair void' extending to a few gravitational radii from the black hole. Pair fractions peak in the midplane just outside the plunging region and within a thin strip at the base of the corona. For moderate to high accretion rate models, pairs are near equilibrium close to the disk midplane, where the scattering optical depth is high and pair equilibrium timescales are short, and could be comparable to the Coulomb collision timescale. This suggests the possibility of a pair-regulated coronal temperature. In contrast, the upper corona and jets, where the scattering optical depth is relatively low and pair equilibrium timescales are long, are populated with pairs that may exceed their equilibrium value by orders of magnitude. These pairs are transported by advection from the disk, which dominates over local pair processes. This result highlights advection as a significant source of pair injection, which may be relevant for certain X-ray binaries exhibiting $\gamma$-ray signatures. The pair density along the magnetically dominated poles exceeds the Goldreich-Julian density in some models.
\end{abstract}
%. .

% Select between one and six entries from the list of approved keywords.
% Don't make up new ones.

%%%%%%%%%%%%%%%%%%%%%%%%%%%%%%%%%%%%%%%%%%%%%%%%%%

%%%%%%%%%%%%%%%%% BODY OF PAPER %%%%%%%%%%%%%%%%%%

%==============================================================================================================%

\section{Introduction} \label{sec:intro}
There is increasing interest in the field of black hole astrophysics to explore whether electron-positron pairs play an important role in accreting black holes. In the optically thin limit, electron-positron pairs are believed to play a critical role in screening the electric field and conducting electric current in a Blandford–Znajek jet \citep{1977MNRAS.179..433B}. The mechanism for supplying the required pair plasma to the force-free jet has been a longstanding issue. One proposed mechanism is the gap acceleration model \citep{2011ApJ...730..123L}, in which an unscreened electric field accelerates electrons to high Lorentz factors. The electrons then emit and Compton upscatter photons, and the collisions of these high-energy photons produce electron-positron pairs. The newly produced pairs undergo the same acceleration, establishing a pair cascade until the electric field is screened. The gap acceleration model has been extensively studied within the framework of both Newtonian and general-relativistic (GR) particle-in-cell simulations \citep{2018ApJ...863L..31C, 2019PhRvL.122c5101P, 2020PhRvL.124n5101C, 2024arXiv241019061V, 2025arXiv250308487Y}. These studies reveal that gaps are formed in a time-variable manner and can supply the jet with sufficient pair plasma to activate the Blandford–Znajek mechanism. 

In contrast, the drizzle model relies on upscattered photons from the accretion disk colliding with each other within the jet. The drizzle model has been investigated through post-processing radiative GR magnetohydrodynamics (GRMHD) simulations \citep{2011ApJ...735....9M, 2021ApJ...907...73W, 2021MNRAS.507.4864Y}, demonstrating that it can provide the jet with a pair density exceeding the Goldreich–Julian limit \citep{1969ApJ...157..869G}, sufficient to screen the electric field for certain parameters of M87, but not Sgr~A*. The impact of pair plasma has also been explored in the context of observables, including its imprint on polarimetric signatures \citep{2021ApJ...923..272E} and the distinct spectral shapes observed in X-ray binaries \citep{2024NatCo..15.7026N}.

The impact of pair physics on the dynamics and thermodynamics of accretion flows remains relatively unexplored. Plasma surrounding accreting black holes is thought to reach relativistic temperatures (i.e., temperature close to the electron rest mass of $0.511$\,MeV), facilitating the production of high-energy radiation. Additionally, power-law X-ray spectra observed in active galactic nuclei (AGN) suggest that most X-ray energy is emitted at the high-energy end and within a compact region \citep{1989IAUS..134..194D, 1989MNRAS.240...81D, 2013ApJ...769L...7R}. Consequently, plasma around accreting black holes serves as an ideal environment for electron-positron pair production to operate. To quantify the likelihood of pair production, it is useful to define the radiative compactness \citep{1983MNRAS.205..593G}:
\begin{equation}
  l_{\rm rad} = \frac{L}{R}\frac{\sigma_{t}}{m_{e}c^{3}}
\end{equation}
where $L$ denotes the source luminosity, $R$ represents the source size, $\sigma_{t}$ is the Thomson cross section, $m_{e}$ is the electron mass, and $c$ is the speed of light. Observations indicate that many AGN exhibit $l_{\rm rad} \geq 1$ and could be as large as $1000$ \citep{1989MNRAS.240...81D, 2015MNRAS.451.4375F, 2017MNRAS.467.2566F}, meaning that they should be radiatively compact enough to produce electron-positron pairs. This has led to pioneering studies on the thermodynamic effects of enforcing pair equilibrium in mildly relativistic thermal plasma \citep{1981ApJ...251..713L, 1982ApJ...258..335S, 1982ApJ...253..842L, 1984MNRAS.209..175S}. It has been demonstrated that, for a given proton scattering optical depth, two series of pair equilibrium solutions exist - one with a high pair density and another with a low pair density. Moreover, given a proton scattering optical depth, which is the electron scattering optical depth in the absence of pairs , there is a maximum temperature beyond which no pair equilibrium solution can be achieved. When this critical temperature is exceeded, pair annihilation always occurs at a slower rate than pair production. The presence of a magnetic field \citep{1983AIPC..101..400T, 1983ApJ...269...49A, 1984ApJ...285..747B, 1999ApJ...517..381E} produces qualitatively similar effects. Consequently, the requirement for pair equilibrium imposes a limit on the maximum allowable electron temperature \citep{1971SvA....15...17B, 1991ApJ...380L..51H, 1993ApJ...413..507H, 1994ApJS...92..555F}.  

In a later series of seminal works incorporating pair physics into analytical models of accretion disks \citep{1988PASJ...40..435K, 1990PASJ...42..347K, 1992ApJ...394..500B, 1994ApJ...422..737K, 1996ApJ...468..330K}, this temperature limitation manifests as the absence of equilibrium solutions within a specific radial range \citep{1987ApJ...313..689B,1989ApJ...340.1024W}, which emerges when the black hole’s mass accretion rate surpasses a critical threshold. Later, the physics of pairs was included in the global structure of geometrically thick, advection-dominated, hot accretion flows \citep{2014ARA&A..52..529Y}. However, these studies find that the global accretion disk structure is not significantly affected by pair production unless the accretion rate approaches the Eddington limit \citep{kusunose1996effects, 1996ApJ...467...99B, 1999ApJ...517..381E, 2020JPhCS1640a2022S}.

Recent observations of the compactness parameter in AGN suggest that most AGN do not exceed the pair equilibrium threshold \citep{2015MNRAS.451.4375F, 2017MNRAS.467.2566F, 2022ApJ...927...42K}, although there are contradictory views on this topic \citep{2017ApJ...847...96L}. This indicates that pair production may play a role in regulating global accretion dynamics by controlling the temperature of the accretion flow through the thermostat mechanism \citep{1993ApJ...413..507H, 1995ApJ...447..526P, 1997ApJ...487..759D, 1997ESASP.382..373Z, 1997ApJ...487..759D, 2021ApJ...914L...5Z}. Specifically, when pairs exceed the temperature defined by the equilibrium line, pair production enters a runaway phase. The newly generated pairs redistribute energy among leptons, enhancing radiative cooling due to an increase in lepton number density, thereby lowering the plasma temperature until pair production ceases \citep{1987ApJ...321..186K, 1989ApJ...340.1024W, 2019BAAS...51c.126K}. To investigate this mechanism, it is necessary to compare the timescales of pair production with those of advection, monitor the spatiotemporal distribution of pairs, and incorporate relevant pair production physics into a global simulation of an accreting black hole. We chose to approach this problem within the framework of 3D GRMHD, as it is computationally less expensive than the general relativistic particle-in-cell method, albeit at the cost of reduced accuracy in radiative and pair production processes while still capturing the global dynamics of accretion flows. 

Note that we are not the first group attempting to incorporate the physics of pair production into a 3D GRMHD simulation. Notably, \citet{2005astro.ph..6368M} and \citet{2005astro.ph..6369M} injected electron-positron pairs using the approximate rates from \citet{1999ApJ...518..356P} and \citet{1999ApJ...524..262M} through the GRMHD equations. However, their model assumed a single-component fluid and primarily focused on studying the composition of the relativistic jets. Additionally, it is not possible to explicitly examine the number density of pairs in a single-component model, especially since the total fluid density remains susceptible to artificial flooring due to the plasma magnetization limit. Here, we adopt a slightly different approach by explicitly advecting and tracking the positron mass density. This approach makes reference to the multi-species, single-fluid MHD model commonly used in the space physics community \citep{2002JGRA..107.1282M, 2009AsBio...9...55T, 2010JGRA..115.4201W, 2013JGRA..118..321M, 10.1115/IMECE2018-87467, 2020JGRA..12526945S, 2021ApJ...921..139L, 2023ApJ...958..142L}, and also the two-fluid model in \citet{2021ApJ...914..138C, 2022ApJ...941..115C, 2023ApJ...945..133C}. This method allows us to monitor the spatiotemporal distribution of pairs without being influenced by artificial mass injection due to the plasma magnetization limit.  

As such, we are interested in the following scientific questions: \textbf{\emph{i})} \textit{What is the pair equilibrium timescale? How does it compare to other relevant timescales?}, \textbf{\emph{ii})} \textit{What is the spatial-temporal distribution of pairs?}, \textbf{\emph{iii})} \textit{Do electron-positron pairs respect the one-zone equilibrium model?}, and \textbf{\emph{iv})} \textit{If not, what prevents them from reaching equilibrium?}

This paper is organized as follows: Section \ref{sec:method} outlines the tools and methods used to simulate global accretion flows while incorporating pair physics. Section \ref{sec:results} presents the major results of this study. Section \ref{sec:imply} discusses the implications and astrophysical significance of our findings. Section \ref{sec:conclu} concludes the study by briefly recapping the major results, listing current limitations, and discussing possible future improvements. In Appendix \ref{app:svensson}, we recap the pair equilibrium curve presented by \citet{1984MNRAS.209..175S}. In Appendix \ref{app:rates}, we review the physics of pair production, and in Appendix \ref{app:radiative}, we outline the radiative processes incorporated.

%==============================================================================================================%

\section{Methodology} \label{sec:method}

%==============================================================================================================%

\subsection{GRMHD Simulations} \label{subsec:grmhd}

To model pair productions in a global accretion flow around black holes, we solve the ideal GRMHD equations using the open-source code \texttt{iharm3d} \citep{2003ApJ...589..444G, 2006ApJ...641..626N, prather2021iharm3d}, where the equations (with speed of light $c = 1$, gravitational constant $G = 1$, and black hole mass $M_{\rm BH}= 1$) are \citep{1989cup..book.....A}
%,%:%
\begin{equation} \label{eqn:grmhd}
\begin{aligned}
    \partial_{t}(\sqrt{-g}\rho u^{t}) &= -\partial_{i}(\sqrt{-g}\rho u^{i}), \\
    \partial_{t}(\sqrt{-g}T^{t}_{\;\;\nu}) &= -\partial_{i}(\sqrt{-g} T^{i}_{\;\;\nu}) + \sqrt{-g}T^{\kappa}_{\;\;\lambda}\Gamma^{\lambda}_{\;\;\nu\kappa} - \sqrt{-g}\Lambda u_{\nu}, \\
    \partial_{t}(\sqrt{-g}B^{i}) &= -\partial_{j}[\sqrt{-g}(b^{j}u^{i} - b^{i}u^{j})], \\
    \partial_{i}(\sqrt{-g}B^{i}) &= 0,
\end{aligned}
\end{equation}
where $g_{\mu\nu}$ is the Kerr metric in the funky modified Kerr-Schild coordinates \citep{prather2022grmhd}, $\sqrt{-g}$ is the determinant of the metric, $\rho$ is the total gas density, $\Gamma$ is the Christoffel symbol, $B^{i}$ is the magnetic field in the coordinate frame, $u^{\mu}$ is the $4-$velocity, $b^{\nu}$ is the magnetic $4-$vector, and the stress-energy tensor is expressed as \citep{dhang2023magnetic}:
\begin{equation}
    T^{\mu\nu} = (\rho h + b^{2})u^{\mu}u^{\nu} + (P + \frac{b^{2}}{2})g^{\mu\nu} - b^{\mu}b^{\nu},
\end{equation}
where $h = 1 + \gamma/(\gamma - 1)P/\rho$ is the specific enthalpy. The $\nabla\cdot \vec{B} = 0$ constraint is maintained by the flux-constrained transport scheme \citep{toth2000b}. We use the $5$th order Weighted-Essential-Non-Oscillatory scheme \citep{jiang1996efficient} to reconstruct primitive variables to the cell boundaries and the Lax-Friedrichs solver to build the Riemann fluxes. We evolve the GRMHD equations using the $2$nd order Strong Stability Preserving Runge-Kutta Method \citep{gottlieb2011strong} with a Courant–Friedrichs–Lewy number of $0.7$.

The accreting plasma is an ideal gas with an adiabatic index of $\gamma = 5/3$, and the initial condition is the Fishbone-Moncrief (FM) torus \citep{1976ApJ...207..962F}. In this study, we consider only the SANE accretion state \citep{2012MNRAS.426.3241N}. The initial torus has an inner radius at $10$ and a pressure maximum at $20$, all in units of $GM/c^{2}$. The small torus size is consistent with our short simulation duration, and to ensure that the torus achieves inflow equilibrium within our region of interest. Additionally, a small torus size prevents accumulation of large scale magnetic fields on the black hole horizon and helps the accretion flow remain SANE. The initial magnetic field within the torus is set via the vector potential \citep{wong2022patoka}:
\begin{equation} \label{eqn:sane}
    A_{\phi} \propto \text{max} \left(\frac{\rho}{\rho_{\rm max}} - 0.2, 0 \right),
\end{equation} 
where $\rho_{\rm max}$ is the maximum density, $r$ is the radial coordinate and $\theta$ is the polar angle. The normalization constant is set so that the maximum gas pressure $(P_{g})_{\rm max}$ and the maximum magnetic pressure $(P_{\rm mag})_{\rm max} = (b^{2}/2)_{\rm max}$ combine to give $(P_{g})_{\rm max}/(P_{\rm mag})_{\rm max} = 100$. The radial outer boundary is at $1,000$ $GM/c^{2}$ with outflowing boundary conditions. The inner boundary is placed such that there are $5$ active zones within the black hole horizon, which is given as $r_{\rm BH} = 1 + \sqrt{1 - a^{2}}$ with $a$ being the spin of the black hole. Matter is allowed to flow into the boundary but not vice versa. Axisymmetric boundary conditions are employed along the pole, while periodic boundary conditions are assumed along the azimuthal direction. 

The FMKS coordinate system is characterized by a set of parameters ($h$, $s$, $B$, $\alpha$), and we choose to follow \citet{wong2022patoka} in adopting the values ($0.3$, $0.5$, $0.82$, $14$). We follow a number of previous works \citep{dexter2020sgr, dexter2020parameter, dexter2021radiation, gupta2024shock} to choose a resolution of $320 \times 256 \times 160$. We verify that our choice of grid setup is adequate to resolve the fastest growing mode of the magnetorotational instability (MRI).  The wavelength of the MRI fastest growing mode \citep{noble2010dependence} can be approximated as:
\begin{equation} 
    \lambda_{\theta} \approx 9.17(\frac{H}{r})r\beta^{-1/2},
\end{equation}
assuming a Keplerian rotating thin disk. Here, $H/r$ is the aspect ratio of the disk, $r$ is the radial coordinate, and $\beta = 2P/b^{2}$ is the plasma beta. The polar grid size is $\Delta h_{\theta} \approx r\Delta\theta$. The quality factor $Q_{\theta} = \lambda_{\theta}/\Delta h_{\theta}$ should be at least $10$. Assuming $H/r = 0.1$ and $\beta = 100$, we find the required polar discretization $\Delta\theta \approx 0.01$. Thus the number of grids required within an aspect ratio of $H/r = 0.1$ is $\approx 10$. We compute the number of cells per scale height averaged with respect to the natural log of the radial coordinate, log$(r)$ \citep{noble2010dependence} by: 
\begin{equation} 
   \langle N \rangle = \frac{\int N(r) d\text{log}(r)}{\int d\text{log}(r)},
\end{equation}
where $N(r)$ is the number of grids within $\pi/2 - H/r \leq \theta \leq \pi/2 + H/r$. The integration is done from the inner boundary to the radius of maximum density. Our evaluation shows that $\langle N \rangle$ for $H/r = 0.1$ is $40$, larger than the resolution requirement. 

%==============================================================================================================%

\subsection{Radiative Cooling} \label{subsec:cooling}

We study the impact of pair production in black holes accreting at different rates. The accretion disk adjusts its scale height accordingly \citep{dexter2021radiation, 2025MNRAS.541.3184H}. Therefore, we need to model accretion disks that attain a specific scale height. To achieve this, we add a cooling rate $\Lambda$ to Equation \ref{eqn:grmhd}. We set $\Lambda$ so as to remove thermal energy generated via grid-scale dissipation and to maintain the accretion disk at a certain aspect ratio $H/r$. We consider the cooling function by \citet{2009ApJ...692..411N, noble2010dependence}, which reads:
\begin{equation}
\begin{aligned}
    \Lambda &= s\Omega u [Y - 1 + |Y - 1|]^{q}, \\
    Y &= \frac{(\gamma - 1)u}{\rho T_{*}}, \\
    T_{*} &= \frac{\pi}{2}\frac{R_{z}(r)}{r}\left[\frac{H}{r} \right]^{2}, 
\end{aligned}
\end{equation}
where $u = P/(\gamma - 1)$ is the gas internal energy and $q = 1/2$. Here, we fix the aspect ratio $H/r$ to be constant everywhere, and its value will be one of the model parameters. We choose $s = 1$. We demand that the maximum change in the conservative variables $U$ due to cooling in each timestep, $\Delta U$, cannot exceed $10$\,\% and we manually replace $\Lambda \rightarrow 0.1\Delta U/\Delta t$, where $\Delta t$ is the timestep, in case this requirement is not satisfied. The relativistic correction factor $R_{z}(r)$ is given as:   
\begin{equation}
    R_{z}(r) = \frac{1}{r}[l_{k}^{2} - a^{2}(\epsilon_{k}^{2} - 1)],
\end{equation}
where $l_{k}$ ($\epsilon_{k}$) is the specific angular momentum (energy) of a circular orbit around a spinning black hole. Also, $\Omega$ is the orbital frequency which is defined as:
\begin{equation} \label{eqn:omega}
\Omega = 
\begin{dcases}
    \frac{1}{r^{3/2} + a}, & \text{if } r \geq r_{\rm isco}\\
    \frac{g^{\phi\mu}(r,\theta=\pi/2)K_{\mu}}{g^{t\mu}(r,\theta=\pi/2)K_{\mu}}, & \text{otherwise}
\end{dcases}
\end{equation}
where $r_{\rm isco}$ is the radius of the innermost stable circular orbit (ISCO, or plunging region) and $K_{\mu}$ is the $4-$velocity of the ISCO orbit, so that:
\begin{equation}
\begin{aligned}
  K^{\mu} &= (K^{t}, 0, 0, K^{\phi}), \\
  \frac{K^{\phi}}{K^{t}} &= \frac{1}{r_{\rm isco}^{3/2} + a}.
\end{aligned}
\end{equation}
and for $r < r_{\rm isco}$, both $l_{k}$ and $\epsilon_{k}$ take their values at the ISCO. Finally, the cooling function is applied only to bound elements, i.e., fluid with $(1 + u/\rho + P/\rho + b^{2}/\rho)u_{t} > -1$. This is to prevent the cooling of the jet. 

%==============================================================================================================%

\subsection{Pair Production} \label{subsec:pair}

To track electron-positron pairs in global accretion simulations, we evolve additionally the total positron mass density $\rho_{+} = n_{+}m_{e}$, where $n_{+}$ is the positron number density and $m_{e}$ is the electron mass, as a passive scalar:
\begin{equation} \label{eqn:pair_advect}
    \partial_{t}(\sqrt{-g}\rho_{+} u^{t}) = - \partial_{i}(\sqrt{-g}\rho_{+} u^{i}) 
\end{equation}
where we assume that protons, electrons, and positrons are advected with a common velocity field $u^{\mu}$. We enforce charge neutrality so that $n_{e} = n_{+} + n_{p}$. Thus, $n_{e}$, $n_{+}$, and $n_{p}$ could be uniquely determined given $\rho_{+}$ and $\rho$. We solve the pair production via operator split:
\begin{equation} \label{eqn:onezone}
\begin{aligned}
    \frac{dn_{+}}{dt} &= \frac{\dot{n}_{+}}{u^{t}}, \\
    \dot{n}_{+} &= \dot{n}_{C} - \dot{n}_{A}.
\end{aligned}
\end{equation}
where $\dot{n}_{C}$ ($\dot{n}_{A}$) is the pair production (annihilation) rate, where formulas are given in Appendix \ref{app:rates}. We also review the radiative processes involved in pair production in Appendix \ref{app:radiative}. The extra division by $u^{t}$ is to convert from coordinate time to proper time. Note that even though we assume the positrons advect passively and do not directly affect the overall dynamics of the plasma, the pair production rate should feed back to the total mass density $\rho$ as $d\rho/dt=2d\rho_{+}/dt$, and thus would indirectly affect the dynamical evolution (Equation \ref{eqn:grmhd}). However, as we shall see, the maximum pair fraction $z = n_{+}/n_{p}$ is $\sim \mathcal{O}(0.01)$ so the maximum pair mass density is $\sim 10^{-5}$ of the proton mass density. We thus expect that the feedback of pair production to the main flow is negligible, at least within our current framework. To compute the pair production rate, one must compute the lepton scattering optical depth $\tau_{T}$ and the (local) scale height $H$ in each grid cell. We estimate $\tau_{T}$ and $H$ as:
\begin{equation} \label{eqn:temperature}
\begin{aligned}
    H &= \frac{c_{s}}{|\Omega|}, \\
    \tau_{T} &= 2H\sigma_{T}(n_{+} + n_{e}), 
\end{aligned}
\end{equation}
where $c_{s} = (\rho/h)(\partial h/\partial \rho)$ is the local (relativistic) speed of sound, $\Omega$ is the angular velocity. To avoid unphysical $\Omega$ in case there are inversion failures or floors applied, we set $\Omega$ to be the minimum between $u^{\phi}/u^{t}$ and $1/(r^{3/2} + a)$, the Keplerian angular velocity. Also, $\sigma_{T}$ is the Thomson scattering cross section. 

%%%%%%%%%%%%%%%%%%%%%%%%%%%%%%%%%%%%%%%%%%%%%%%%%%
%The factor of $2$ accounts for a two-sided disk. 
%%%%%%%%%%%%%%%%%%%%%%%%%%%%%%%%%%%%%%%%%%%%%%%%%%

We remark on some technical concerns in implementing pair production in global accretion simulations. \textbf{First}, the production rate $\dot{n}_{+}$ could be stiff, especially at the first few time steps when we start adding pairs to the simulation domain or if the scattering optical depth of the fluid parcel suddenly gets high. We define the quality factor for the pair production as:
\begin{equation}
    Q_{+} = \frac{n_{+}}{|\dot{n}_{+}|},
\end{equation}
and we require that $Dt \leq \alpha Q_{+}$ where $Dt$ is the time-step of a Runge-Kutta sub-stage, and we choose $\alpha = 0.1$. If this condition does not hold, we will implicitly solve Equation \ref{eqn:onezone} using the Crank–Nicolson method. Otherwise, we solve Equation \ref{eqn:onezone} with the 2nd order Runge-Kutta Method.

\textbf{Second}, we find that if the proton scattering optical depth $\tau_{p} = 2H\sigma_{t}n_{p}$ is larger than some critical $\tau_{\rm crit}$, pair production becomes a runaway process, creating an unphysically large amount of positrons. This occurs due to the absence of a pair equilibrium solution, as mentioned in Section \ref{sec:intro}, and this happens in the densest part of the disk when the accretion rate is approaching the Eddington limit. This is also an artifact of our constant electron temperature assumption, where we do not allow electrons to cool. To tackle this, we choose to turn off pair production but allow pair annihilation (i.e., setting $\dot{n}_{+} = -\dot{n}_{A}$) whenever $\tau_{p} > \tau_{\rm crit}$. 

Our strategy thus implicitly rejects the pair-dominated branch solution ($z \gg 1$), as identified in some previous analytical studies on the impact of pair production in accretion disks \citep{1988PASJ...40..435K, 1990PASJ...42..347K, 1992ApJ...394..500B, 1994ApJ...422..737K, 1996ApJ...468..330K}. This choice is well justified, primarily because the pair-dominated disk is unstable and nonphysical, exhibiting a proton temperature exceeding the virial value and a disk scale height $H/r > 1$ \citep{1988PASJ...40..435K, 1990PASJ...42..347K}. Additionally, studies suggest that accreting black holes may not be pair dominated \citep{1997ApJ...476..620H, 1997ASPC..121..101M, 2017ApJ...847...96L, 2017MNRAS.468.3489K}. Moreover, within the main body of the disk - where the effective optical depth is high - the electron temperature rapidly cools to $\sim 10^{6} - 10^{7}$\,K due to lepton-lepton and lepton-proton Coulomb collisions, as well as the dramatic increase in radiative cooling rate (e.g., thermal bremsstrahlung, which scales with the square of the number density). At such temperatures, pair production is inefficient. Any excess pairs transported to the optically thick disk are annihilated. And as we shall see, the annihilation timescale is short in the dense disk.

\textbf{Third}, in the strongly magnetized region, $n_{p}$ increases due to the artificial addition of proton mass, where density floors are applied. Although the scattering optical depth in the $\sigma = b^{2}/\rho > 1$ region is small, such that the dominant pair production channel is the particle-particle collision, the uncertainty in the actual value of $n_{p}$ might lead to a systematic overestimation (underestimation) of the total pair production rate if $z < z_{\rm eq}$ ($z > z_{\rm eq}$). This is because $n_{p}$ explicitly contributes to the particle-particle pair creation and annihilation rates. To avoid such confusion, we set $n_{p} = 0$ when calculating the pair production rate in the $\sigma > 1$ region. 

%#####################################################################################################################%

\begin{table*}
    \centering
    \caption{List of model parameters. The model with $H/r = 0.3$ is a pure hot accretion flow without cooling, where we find that the density scale height is approximately $0.3$. All models have $M_{\rm BH} = 10^{8}$\,$M_{\odot}$, except for model hr01\_m10 where $M_{\rm BH} = 10$\,$M_{\odot}$. Model hr01 and hr01\_high have $\tau_{\rm crit} = 3.1$, while model hr01\_m10 has $\tau_{\rm crit} = 3.7$. Note that $\tau_{\rm crit}$ is not required for models with $H/r = 0.2$, $0.3$. The series of models with $H/r = 0.1$ exhibits the highest values of $z$, on the order of $\sim \mathcal{O}(0.01)$.}
    \label{tab:params}
    \begin{tabular}{ccccccccc} % four columns, alignment for each
        \hline
        Model & $H/r$ & $\eta$ & $M_{\rm unit}$ & $T_{e}$ (K) & $(\tau_{p})_{\max}$ & $z_{\rm max}/(z_{\rm eq})_{\rm max}$ & $(z/z_{\rm eq})_{\rm max}$ & $(n_{+})_{\rm max}$ (cgs) \\
        \hline
        adaf & $0.3$ & $5\times10^{-5}$ & $1.907\times10^{23}$ & $10^{10.5}$ & $3.069\times10^{-3}$ & $3.475\times10^{-3}$ & $3.475\times10^{-3}$ & $2.571\times10^{2}$ \\
        hr02 & $0.2$ & $5\times10^{-3}$ & $7.575\times10^{24}$ & $10^{9.5}$ & $5.779\times10^{-1}$ & $8.458\times10^{-1}$ & $8.721\times10^{-1}$ & $1.382\times10^{4}$ \\
        hr02\_hot & $0.2$ & $5\times10^{-3}$ & $7.575\times10^{24}$ & $10^{10}$ & $5.779\times10^{-1}$ & $9.055\times10^{-1}$ & $1.944\times10^{0}$ & $6.460\times10^{6}$ \\
        hr01 & $0.1$ & $5\times10^{-2}$ & $7.696\times10^{25}$ & $10^{9}$ & $1.434\times10^{1}$ & $2.357\times10^{-1}$ & $7.729\times10^{3}$ & $1.009\times10^{9}$ \\
        hr01\_m10 & $0.1$ & $5\times10^{-2}$ & $7.696\times10^{11}$ & $10^{9}$ & $1.434\times10^{1}$ & $2.230\times10^{-1}$ & $2.501\times10^{4}$ & $2.641\times10^{16}$ \\
        hr01\_high & $0.1$ & $4\times10^{-1}$ & $6.157\times10^{26}$ & $10^{9}$ & $1.147\times10^{2}$ & $1.440\times10^{-1}$ & $2.721\times10^{6}$ & $5.537\times10^{8}$ \\ 
        \hline
    \end{tabular}
\end{table*}

%#####################################################################################################################%

Our simulations span $H/r = 0.1, 0.2$, and a pure hot accretion flow without cooling (labeled as $H/r = 0.3$). The black hole mass is fixed as $M_{\rm BH} = 10^{8}\,M_{\odot}$ with a spin of $a = +0.9375$. The electron (and positron) temperature $T_{e}$ is a constant, spanning $T_{e} = 10^{9}$, $10^{9.5}$, $10^{10}$, and $10^{10.5}$ K. We also include a model with $H/r = 0.1$, $a = +0.9375$, but with $M_{\rm BH} = 10\,M_{\odot}$ to test our results against black hole mass. The global accretion dynamics in GRMHD simulations is scale free, but the pair production is not. As such, we need to assign the simulation a mass unit $M_{\rm unit}$, a length unit $L_{\rm unit}$, and a time unit $T_{\rm unit}$. The latter two are straightforward since $L_{\rm unit} = GM_{\rm BH}/c^{2}$ and $T_{\rm unit} = GM_{\rm BH}/c^{3}$. We choose $M_{\rm unit}$ by requiring that the mass accretion rate $\dot{M}$:
\begin{equation}
    \dot{M} = -\int\int u^{r}\rho\sqrt{-g}d\theta d\phi,
\end{equation}
as measured at the event horizon be a fraction $\eta$ of the Eddington limit $\dot{M}_{\rm Edd}$, where $L_{\rm Edd} = \dot{M}_{\rm Edd}c^{2}$ and  $L_{\rm Edd} = 4\pi G M_{\rm BH}m_{p}c/\sigma_{t}$ is the Eddington luminosity. To determine a reasonable value of $\eta$ for the accretion disk model, we refer to Table 1 of \citet{dexter2021radiation}. We list all relevant model parameters in Table \ref{tab:params} for reference. Note that only models with $H/r = 0.1$ are sufficiently optically thick to scattering so that we need to impose a limit of $\tau_{\rm crit}$.

Our simulations proceed as follows. First, we run an FM torus in the SANE state for a duration of $10000$\,$GM/c^{3}$. Then, we turn on the cooling function for another duration $10000$\,$GM/c^{3}$. Note that we run the pure hot accretion flow model for the same duration without cooling. Finally, we turn on pair processes and let the disk evolve for an additional $6000$\,$GM/c^{3}$. We adopt a `floor' on $\rho_{+}$ because grid-based hydrodynamics codes cannot handle zeros. In particular, we set a geometrically dependent floor of $\rho_{+,f}$: 
\begin{equation}
    \rho_{+,f} = f\textrm{max}(10^{-6}\frac{r^{-2}}{1 + r/10}, 10^{-20})
\end{equation}
which takes the same form as the geometrical floor of $\rho$, but with $f = 10^{-30}$ to take account of the small value of $n_{+}$. We do not limit $\rho_{+}$ based on other fluid variables because $\rho_{+}$ does not affect the dynamics. At the beginning of the pair injection, we set $z = 10^{-30}$.  

%==============================================================================================================%

\section{Results} \label{sec:results}

%==============================================================================================================%

\subsection{Timescale Analysis} \label{subsec:tscale}
%==============================================================================================================%
%#####################################################################################################################%
\begin{figure*}
    \centering
    \gridline{
    \fig{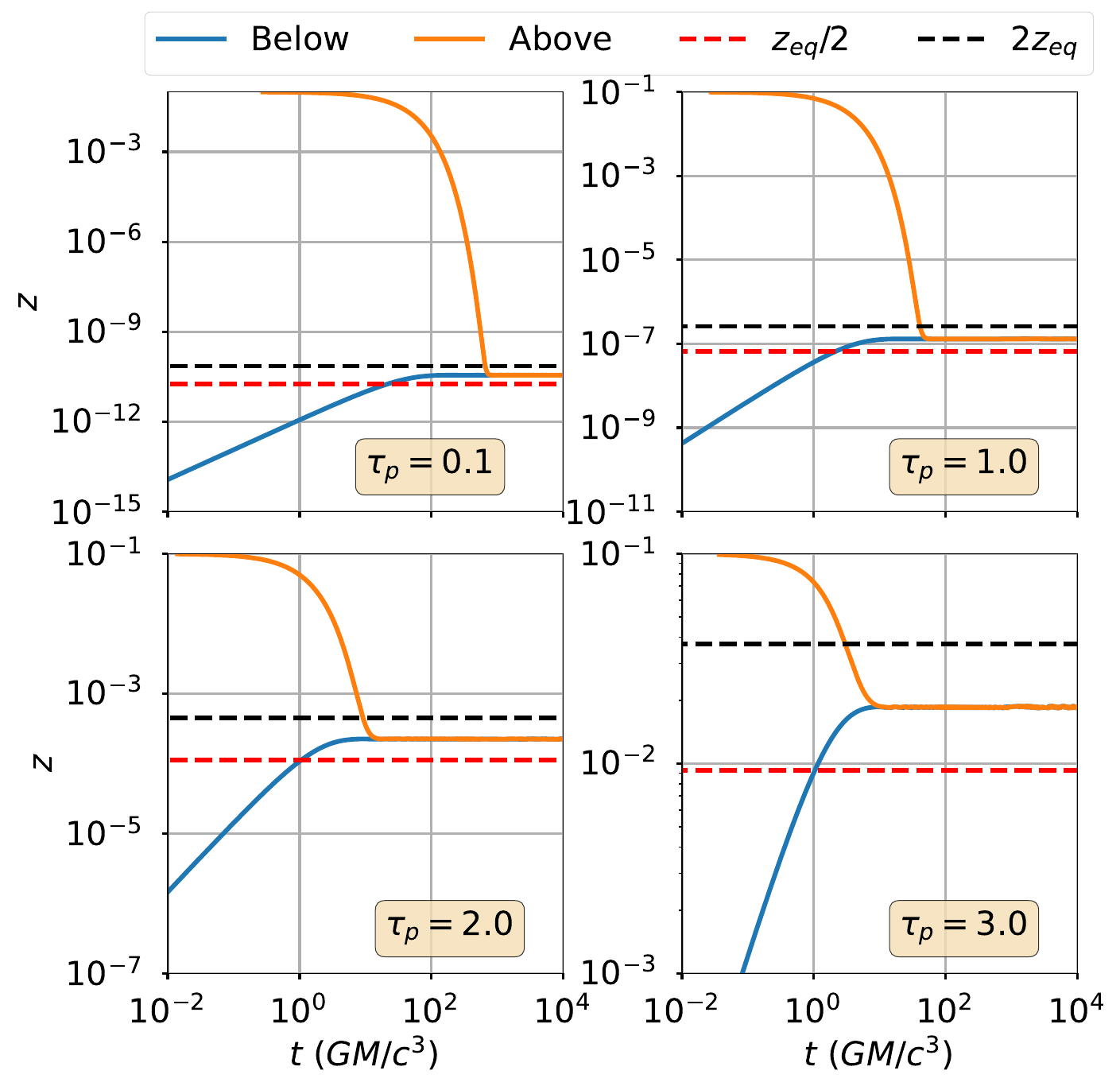}{0.44\textwidth}{(a)}
    \fig{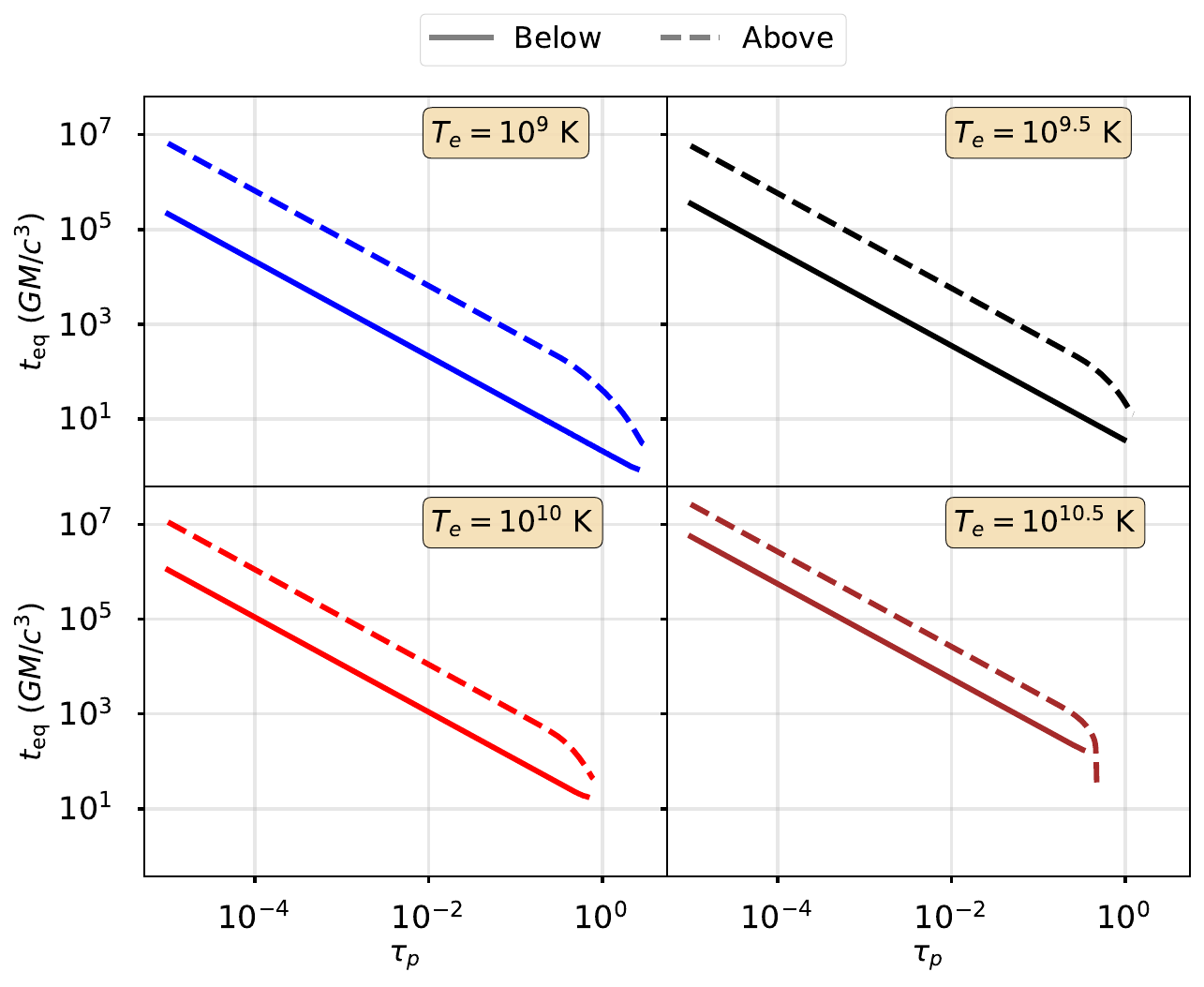}{0.52\textwidth}{(b)}}
    \caption{(a) The time evolution (in $GM/c^{3}$) of $z$ obtained by solving Equation \ref{eqn:onezone}. Here, $T_{e} = 10^{9}$ K and $H = GM_{\rm BH}/c^{2}$ with $M_{\rm BH} = 10^{8}$\,$M_{\odot}$. We show cases of different $\tau_{p} = H\sigma_{t}n_{p}$. Blue (orange) lines indicate initial condition of $z$ being below (above) the equilibrium value $z_{\rm eq}$. We also mark horizontal dashed lines of $2z_{\rm eq}$ (black) and $z_{\rm eq}/2$ (red). The time-coordinate of the intersection between the orange and black (blue and red) is defined as the equilibrium timescale with initial $z$ being above (below) $z_{\rm eq}$; (b) Pair equilibrium timescale against proton scattering optical depth, determined via solving Equation \ref{eqn:onezone} and using the strategy outlined in Figure (a). Here, $\tau_{p}$ ranges from $10^{-5}$ to the point where pair equilibrium is unattainable (i.e., we find $z$ diverges in solving Equation \ref{eqn:onezone}), and $T_{e} = 10^{9}$, $10^{9.5}$, $10^{10}$, and $10^{10.5}$ K. Note that the dashed (solid) line displays the timescale for cases where the initial $z$ is above (below) the equilibrium value.  \label{fig:pair_onezone_svensson}}
\end{figure*}
%#####################################################################################################################%

We briefly review the results of the one-zone equilibrium model presented in \citet{1984MNRAS.209..175S}, focusing on solving Equation \ref{eqn:onezone} until $n_{+}$ reaches pair equilibrium. We examine cases where the initial pair fraction is above or below the equilibrium value $z_{\rm eq}$. We define a pair equilibrium timescale as the time when the pair fraction is more than $0.5$ (less than $2$) of the equilibrium value if the initial value of $z$ is below (above) $z_{\rm eq}$. The initial $z = 10^{-30}$ (below) or $z = 0.1$ (above). We define $H = GM_{\rm BH}/c^{2}$ as the characteristic length scale of the black hole. The proton number density is given by $\tau_{p} \approx H\sigma_{t}n_{p}$. Note that we omit a factor of $2$ here, assuming a spherical cloud of size $H$. Note also that $M_{\rm BH}$ and $\tau_{p}$ are free parameters. 

We show examples of the time evolution of $z$ for fixed $T_{e}$ and $M_{\rm BH}$ but varying $\tau_{p}$ in Figure \ref{fig:pair_onezone_svensson} (a). When $z$ is above (below) $z_{\rm eq}$ at a fixed $T_{e}$, pair annihilation is faster (slower) than creation, and thus $z$ will decrease (increase) towards $z_{\rm eq}$. Note also that using the above definition, the time required to reach equilibrium when $z$ is above (below) $z_{\rm eq}$ is longer (shorter). This is because when the initial $z$ is greater (less) than $z_{\rm eq}$, its time evolution follows an exponential decay (increases linearly). Consequently, the timescale for reaching equilibrium, as per our definition, when the initial $z$ exceeds $z_{\rm eq}$ is given by $t_{\rm above} \sim t_{\rm below} \text{ln}(z_{\rm ini}/z_{\rm eq})$.

We sample values of $\tau_{p}$ from $10^{-5}$ to the value for which no pair equilibrium solution could be found, and vary the electron temperature, spanning $T_{e} = 10^{9}$, $10^{9.5}$, $10^{10}$, and $10^{10.5}$ K. The results are shown in Figure \ref{fig:pair_onezone_svensson} (b). While the equilibrium timescale spans a wide range of values, it tends to approach $10$\,$GM_{\rm BH}/c^{3}$ as $\tau_{p}$ increases towards $1$. For $T_{e} = 10^{9}$ K, the equilibrium timescale could be of order $\mathcal{O}(1)$\,$GM_{\rm BH}/c^{3}$. 

%==============================================================================================================%

\subsection{Model Overview} \label{subsec:summary}
%==============================================================================================================%
%#####################################################################################################################%
\begin{figure*}
    \centering
    \gridline{
    \fig{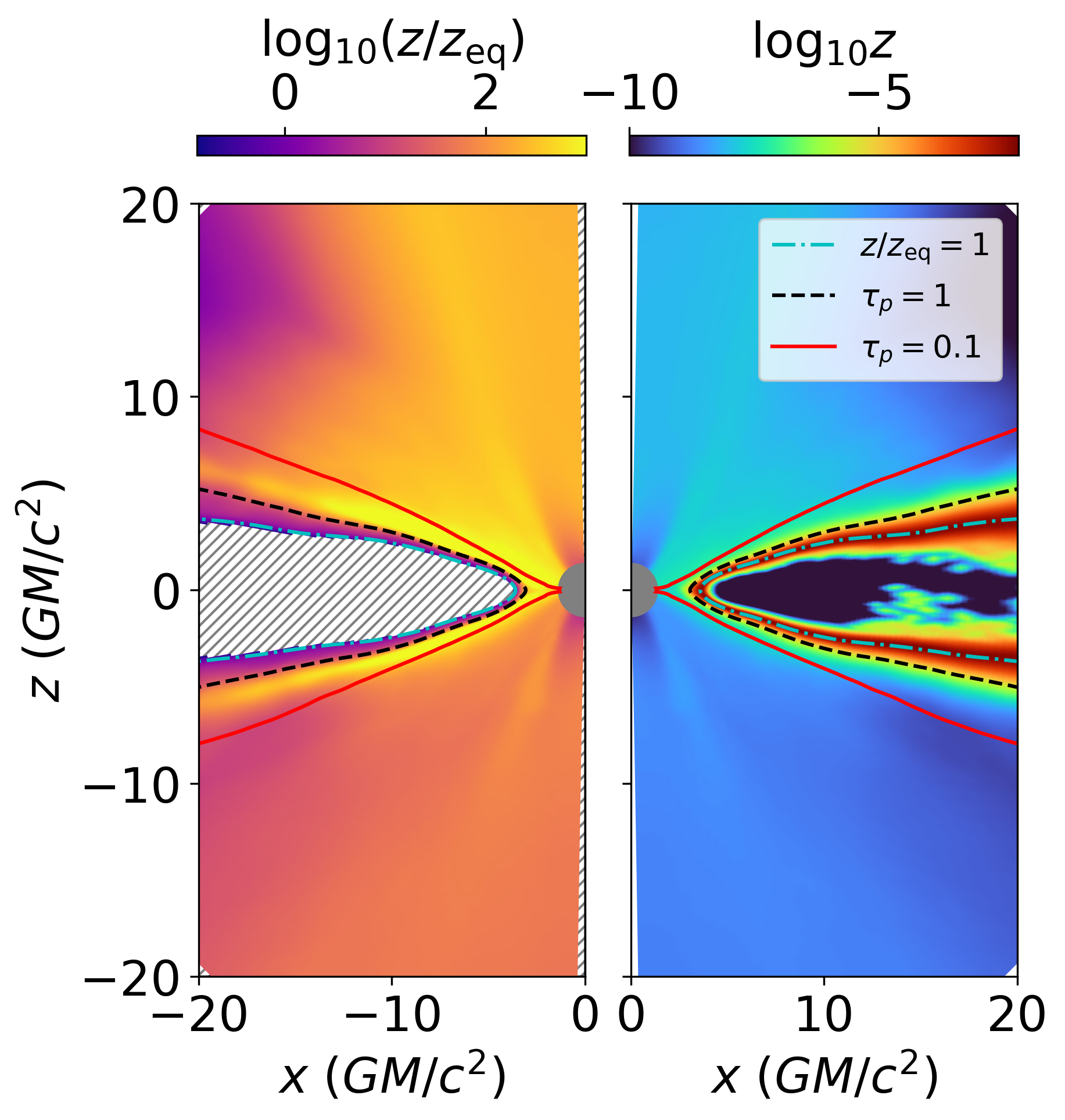}{0.48\textwidth}{(a)}
    \fig{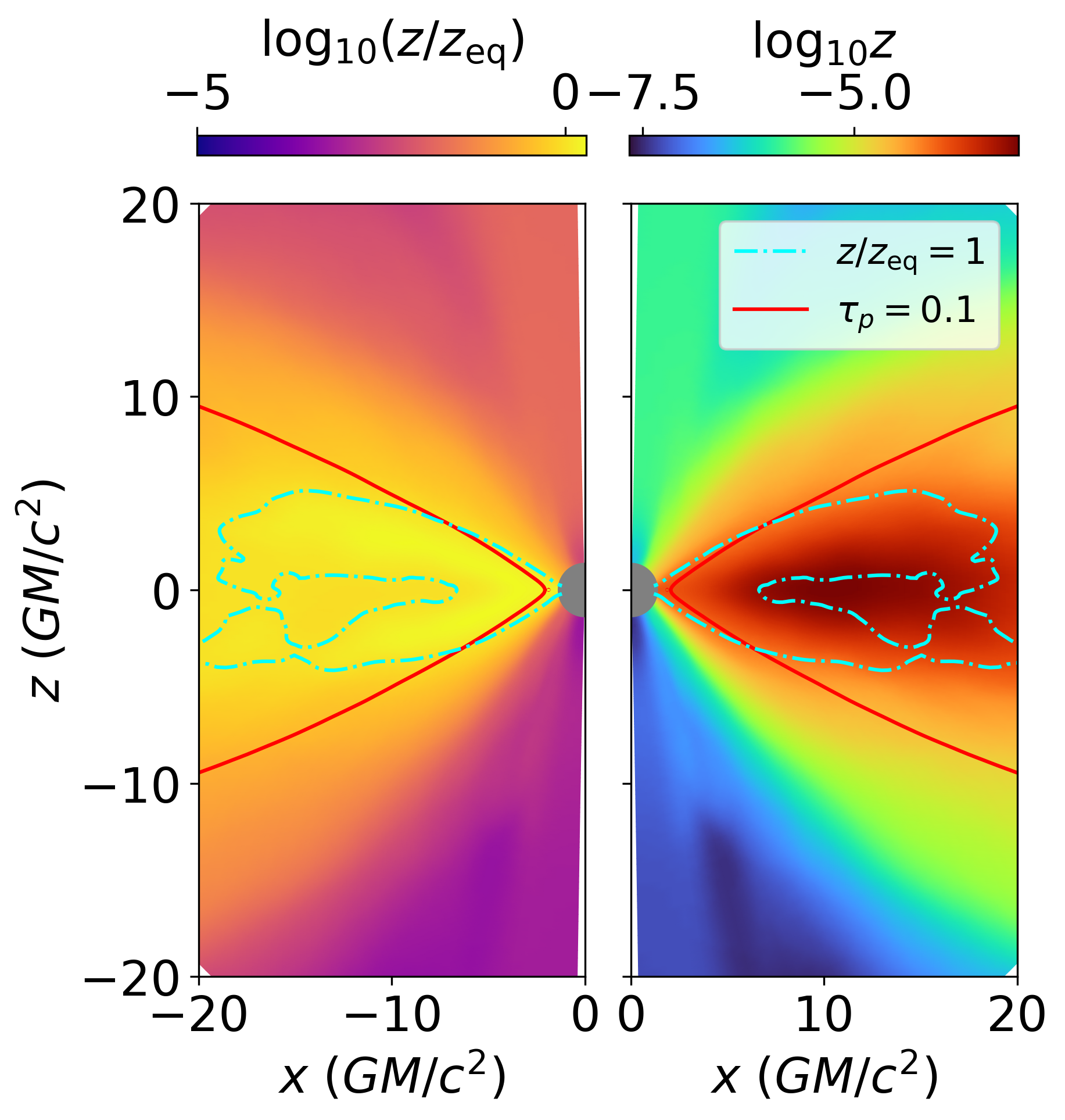}{0.48\textwidth}{(b)}}
    \caption{(a) Spatial distribution of the time- and $\phi$-averaged pair fraction, $z$ (right), and the ratio $z/z_{\rm eq}$ (left), for model hr01, characterized by $\eta = 0.05$ and $T_{e} = 10^{9}$\,$\mathrm{K}$. The equilibrium value $z_{\rm eq}$ is computed using the one-zone equilibrium model, based on the local $\tau_{p}$ derived from time- and $\phi$-averaged fluid variables. Time averaging is performed over the final $2000$\,$GM/c^{3}$ of the pair injection phase, with a cadence of $10$\,$GM/c^{3}$. Contour lines indicate $z/z_{\rm eq} = 1$, $\tau_{p} = 1$, and $\tau_{p} = 0.1$. The hatched region marks zones where $\tau_{p} \geq \tau_{\rm crit}$, and no pair equilibrium solution is available under the constant-temperature assumption. (b) Same as (a), but for model hr02\_hot, with $\eta = 0.005$ and $T_{e} = 10^{10}$\,$\mathrm{K}$. Contour lines for $\tau_{p} = 1$ are omitted. \label{fig:zfrac_zavg}}  
\end{figure*}
%#####################################################################################################################%

%%%%%%%%%%%%%%%%%%%%%%%%%%%%%%%%%%%%%%%%%%%%%%%%%%%%%%%%%%%%%%%%%%%%%%%%%%%%%%%%%%%%%%%%%%%%%%%%%%%%%%%%%%%%%%%%%%%%%%%%%%%%%%
%Before proceeding, we performed a sanity check on our implementation of the pair production scheme. More precisely, we compare the pair annihilation luminosity, $L_{\pm}$, with the black hole luminosity. Here, we define the pair annihilation luminosity as: 
%
%\begin{equation}
%    L_{\pm} = \int2m_{e}c^{2}\dot{n}_{A}\sqrt{-g}dx^{1}dx^{2}dx^{3}, 
%\end{equation}
%
%where the factor of $2$ accounts for the fact that each annihilated positron produces two photons. We find that $L_{\pm}/L$ is on the order of $\mathcal{O}(10^{-2})$ for the high accretion rate model, whereas for the remaining models, this ratio is significantly lower. This suggests that, in our model, the electromagnetic emission due to pair annihilation is subdominant compared to the total plasma emission. 
%%%%%%%%%%%%%%%%%%%%%%%%%%%%%%%%%%%%%%%%%%%%%%%%%%%%%%%%%%%%%%%%%%%%%%%%%%%%%%%%%%%%%%%%%%%%%%%%%%%%%%%%%%%%%%%%%%%%%%%%%%%%%%
 
%. 
%#####################################################################################################################%
\begin{figure*}
    \centering
    \centering
    \gridline{
    \fig{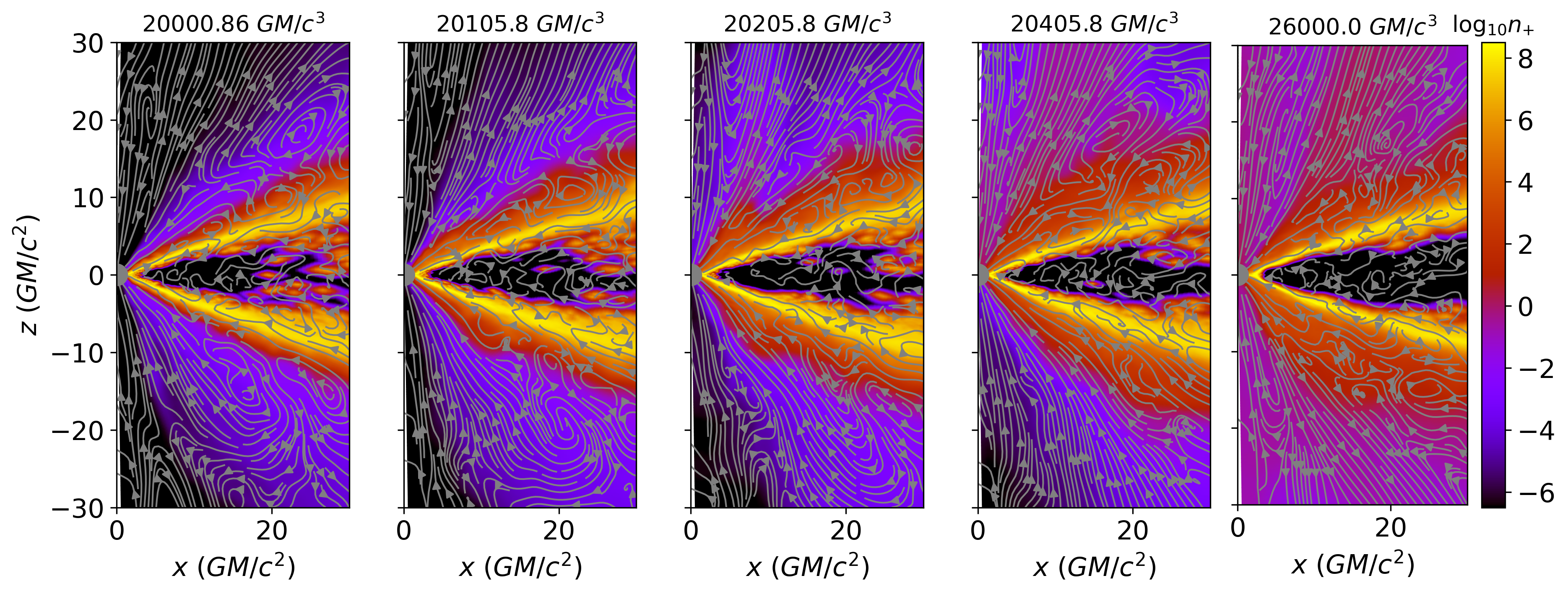}{1.0\textwidth}{(a)}}
    \gridline{
    \fig{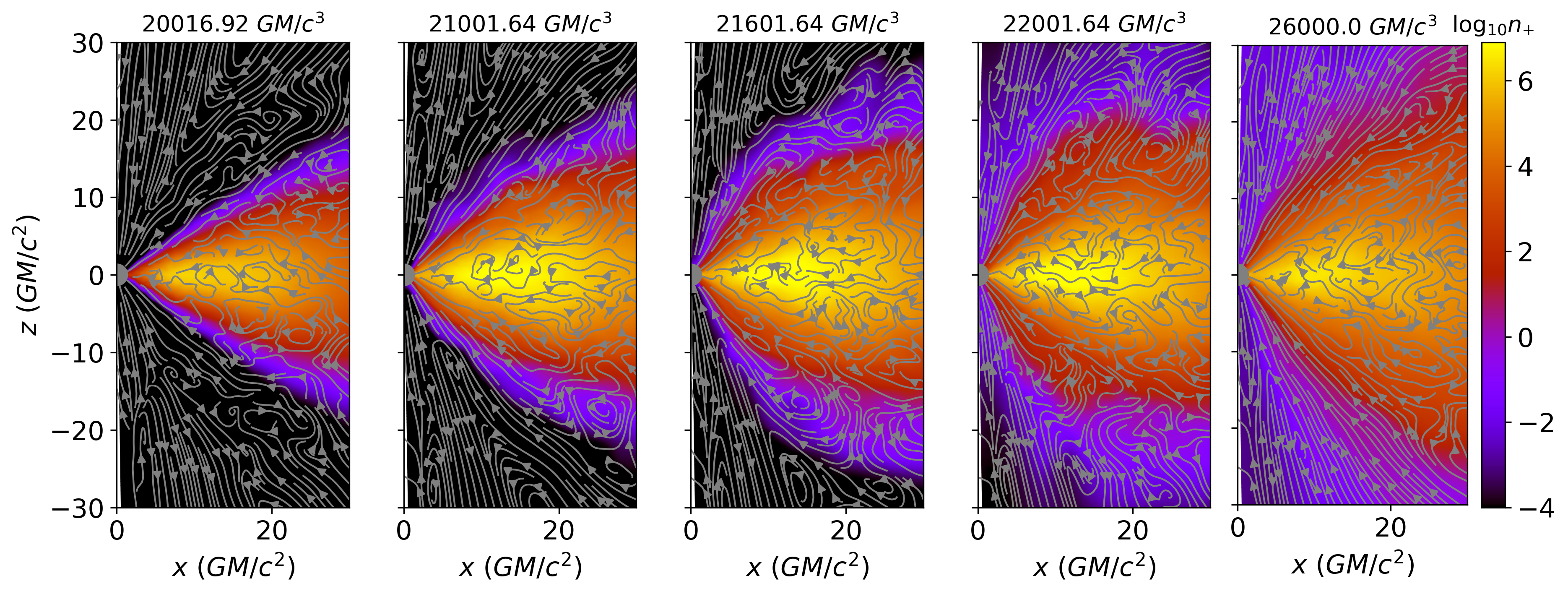}{1.0\textwidth}{(b)}}
    \caption{Positron number density (log$_{10}$ scale) at the onset of the pair injection phase for models (a) hr01\_high ($\eta = 0.4$ and $T_{e} = 10^{9}$\,$\mathrm{K}$) and (b) hr02\_hot $\eta = 0.005$ and $T_{e} = 10^{10}$\,$\mathrm{K}$). Velocity field lines are overlaid in both plots, and all panels within each subplot share a consistent color scale. In (a), pairs rapidly reach their peak and settle into equilibrium within a thin strip at the coronal base, sustained by a short equilibrium timescale. The bulk fluid flow subsequently advects these pairs into the jet and upper corona, where the pair equilibrium timescale exceeds the dynamical timescale. In (b), the midplane pair density rises more gradually due to a longer equilibrium timescale relative to (a), though advection similarly transports pairs into the jet and upper corona. \label{fig:pair_advect_tseries}}
\end{figure*}
\begin{figure}
    \centering
    \includegraphics[width=1.0\linewidth]{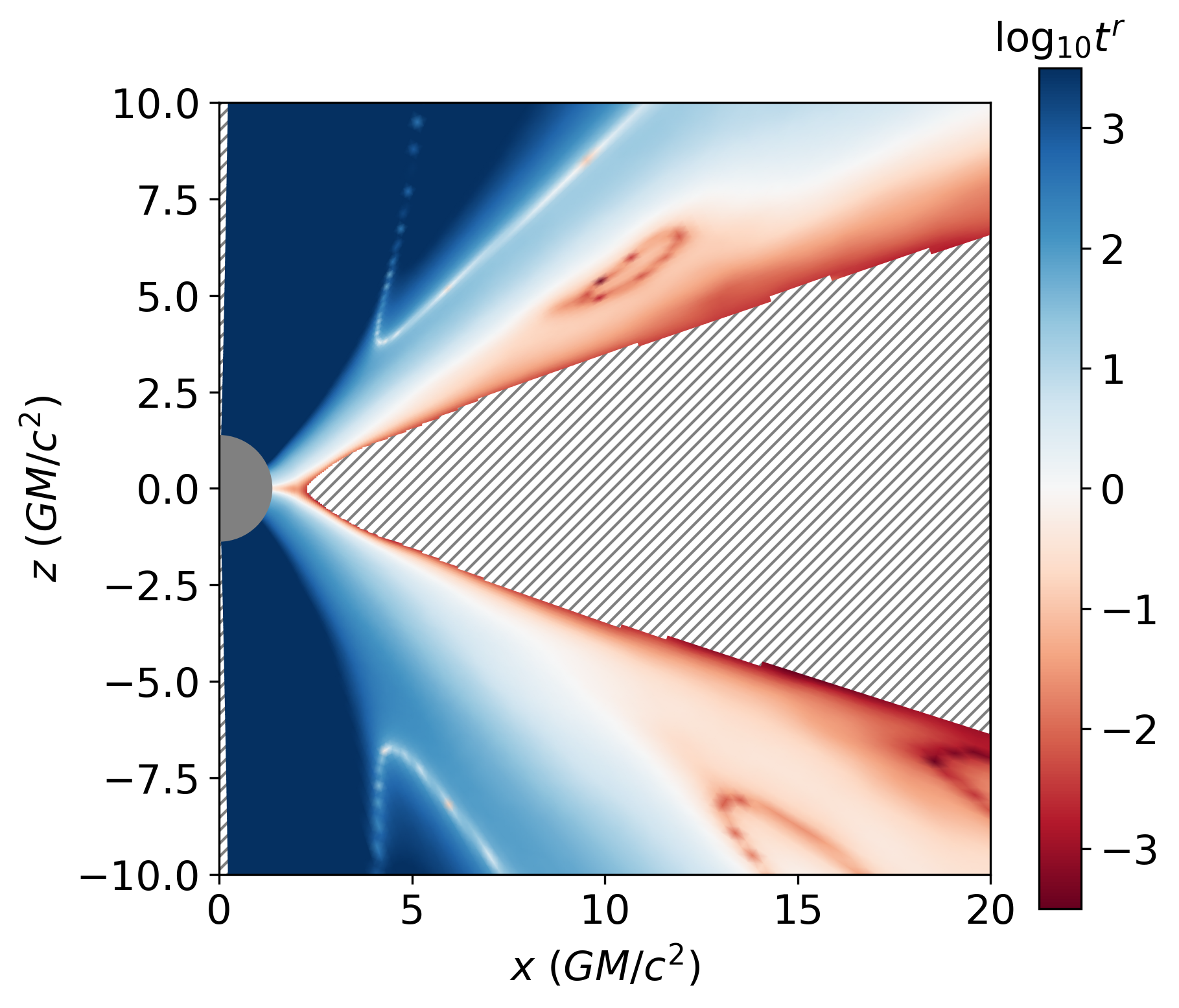}
    \caption{Ratios of the pair equilibrium timescale (computed for an initial $z$ that is above $z_{\rm eq}$) to the inflow timescale, $t^{r}$, for the high accretion rate model hr01\_high. Hatched region has no pair equilibrium solution. \label{fig:compare_tscale}}
\end{figure}
\begin{figure*}
    \centering
    \includegraphics[width=1.0\linewidth]{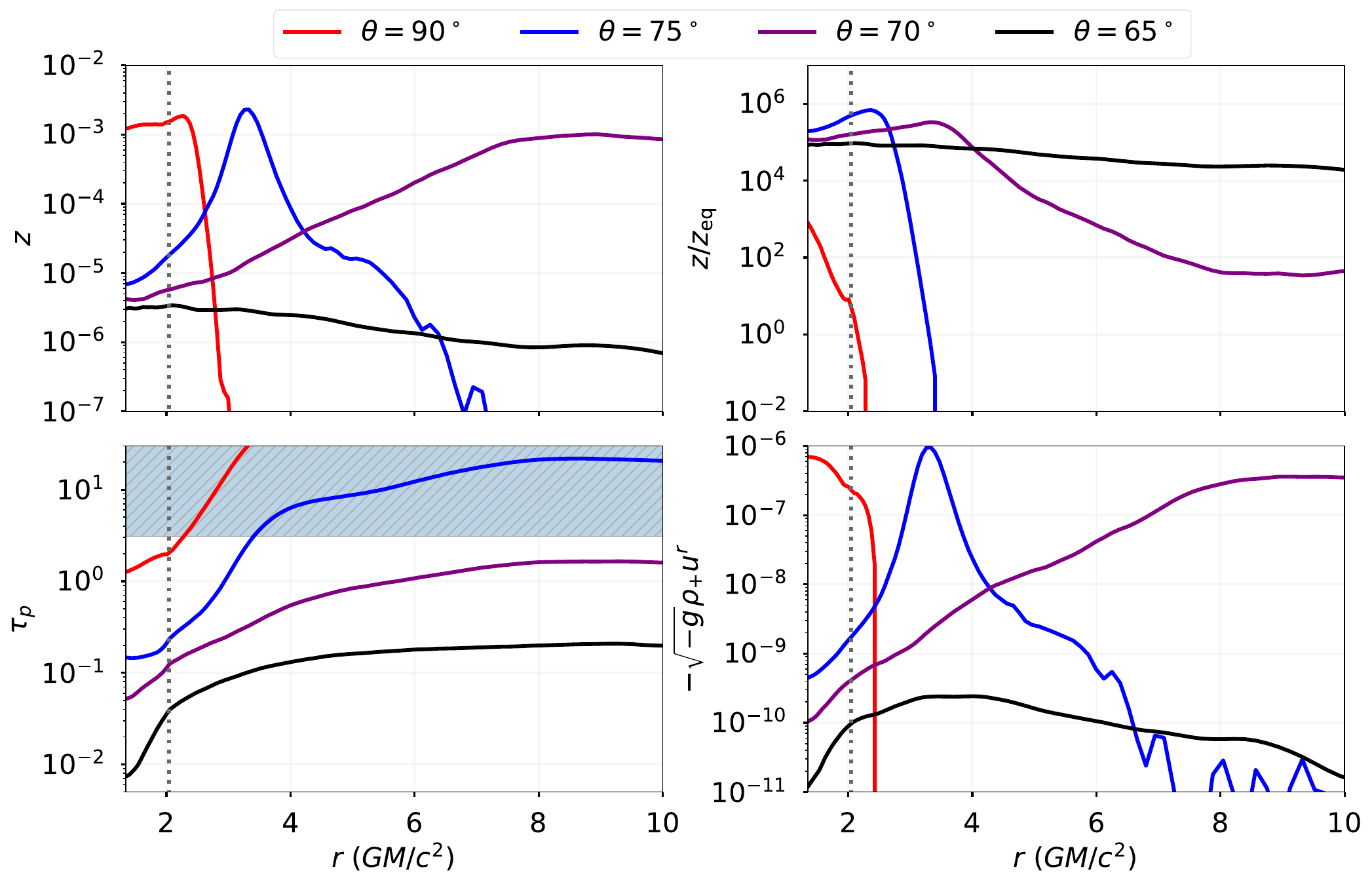}
    \caption{The radial profiles of $z$ (top left), $z/z_{\rm eq}$ (top right), $\tau_{p}$ (bottom left), and the positron mass flux $-\sqrt{-g}\rho_{+}u^{r}$ (bottom right) shown at different altitudes $\theta$ for the high accretion rate model hr01\_high. In the bottom left panel, we shade the region where $\tau_{p} \geq \tau_{\rm crit}$, for which no pair equilibrium solution exists. Note also that the positron mass flux is calculated in code units. Vertical dotted grey lines represent the position of the ISCO. \label{fig:profiles_hr01}}
\end{figure*}
%#####################################################################################################################%

Table \ref{tab:params} summarizes the input parameters and simulation results. We begin by comparing the maximum pair fraction across the simulation domain, $z_{\rm max}$, with the value predicted by the one-zone equilibrium model, $(z_{\rm eq})_{\rm max}$. For the lowest accretion rate model, adaf, $z_{\rm max}$ is three orders of magnitude lower than $(z_{\rm eq})_{\rm max}$. Furthermore, the maximum ratio $(z/z_{\rm eq})_{\rm max}$ is on the order of $\mathcal{O}(10^{-3})$, indicating that the plasma in model adaf is far from pair equilibrium. This deviation arises from the low proton scattering optical depth, $(\tau_{p})_{\rm max} \sim \mathcal{O}(10^{-3})$, which implies that the timescale required to reach pair equilibrium exceeds $10^{5}$\,$GM/c^{3}$ (see Figure \ref{fig:pair_onezone_svensson}(b)). This timescale is significantly longer than other relevant dynamical timescales, suggesting that local pair processes are strongly influenced by global accretion dynamics.

Models with moderate accretion rates - hr02 and hr02\_hot - exhibit moderate proton scattering optical depths but differ in temperature. In both cases, the ratio $z_{\rm max}/(z_{\rm eq})_{\rm max} \approx 1$; High accretion rate models - hr01 and hr01\_10m - have the highest proton scattering optical depths but differ in black hole mass. Their respective maximum pair fractions are $z_{\rm max} = 7.711 \times 10^{-3}$ and $1.726 \times 10^{-2}$, values close to $\mathcal{O}(0.01)$ and only a factor of $\sim 4\text{–}5$ below $(z_{\rm eq})_{\rm max}$. These results indicate that pair equilibrium is nearly attained throughout the domain in both moderate and high accretion rate models Additionally, none of the models considered in this study exhibit $z_{\rm max} \gg (z_{\rm eq})_{\rm max}$, reinforcing the validity of the one-zone pair equilibrium model across the explored parameter space.

Additionally, when all other parameters are held fixed, we find that a higher electron temperature ($T_{e}$; hr02\_hot vs. hr02) and a smaller black hole mass (hr01\_10m vs. hr01) - and thus a smaller characteristic size - lead to a higher maximum positron number density, $(n_{+})_{\rm max}$. Interestingly, the high accretion rate models exhibit $(z/z_{\rm eq})_{\rm max} \gg 1$, suggesting that local pair populations can deviate from equilibrium by orders of magnitude. This highlights the potential for significant pair disequilibrium.
%.   
%==============================================================================================================%

\subsection{Pair Equilibrium and Disequilibrium} \label{subsec:equb}
%==============================================================================================================%

Figure \ref{fig:zfrac_zavg} (a) shows the spatial distribution of $z$ and $z/z_{\rm eq}$ for the high accretion rate model hr01. In this model, $\tau_{p}$ exceeds the critical threshold $\tau_{\rm crit}$ across a broad region near the midplane. Recall that we have turned off pair-production in this region and that this is physically justified by our assumption that a cold, optically thick disk would form (see Section \ref{subsec:coulumb}). As such, pair production is suppressed while only pair annihilation is permitted, causing any existing pairs in the midplane to be annihilated and keeping $z$ small. This `pair void' extends several $GM/c^{2}$ from the black hole horizon.

Above the pair void, a thin strip at the base of the corona exhibits the highest values of $z$. This region coincides with the contour lines $\tau_{p} = 1$ and $z/z_{\rm eq} = 1$, indicating that $\tau_{p}$ is sufficiently close to the critical threshold $\tau_{\rm crit}$. The strip remains near pair equilibrium and produces a substantial number of pairs. As $\theta \rightarrow 0$ or $\pi$ away from the midplane, $\tau_{p}$ gradually decreases to $0.1$. At these lower proton scattering optical depths, $z$ drops by several orders of magnitude, reflecting the sharp decline in pair production rate with decreasing $\tau_{p}$. Note that our discussion of the corona implicitly assumes a sandwich geometry \citep{2010ApJ...712..908S}.

In this representative high accretion rate model, while $z$ decreases as $\theta \rightarrow 0$ or $\pi$, the ratio $z/z_{\rm eq}$ does not. Instead, $z/z_{\rm eq} = 1$ is confined to a thin strip at the coronal base and increases as $\theta$ approaches the poles. Across much of the upper corona, $z/z_{\rm eq}$ can reach values on the order of $\mathcal{O}(10^{2})$. However, it is important to note that in these regions of elevated $z/z_{\rm eq}$, the absolute value of $z$ remains low. Consequently, pairs are not being generated in excess of the maximum predicted by the one-zone pair equilibrium model, $(z_{\rm eq})_{\rm max}$, which is evaluated by setting $\tau_{p} = \tau_{\rm crit}$. Rather, these pairs originate from the thin strip at the coronal base. Advection dominates over local pair processes and acts as the primary mechanism for transporting pairs from this strip into the upper corona. Similar behavior is observed in the other high accretion rate models.

The moderate accretion rate model hr02\_hot also exhibits $(z/z_{\rm eq})_{\rm max} > 1$, though less prominently than the high accretion rate models. Its spatial distribution is shown in Figure \ref{fig:zfrac_zavg} (b). In this model, $\tau_{p}$ never exceeds its critical value, so no pair void is present. The maximum values of $z$ occur along the midplane, where $\tau_{p}$ is highest across the domain, and decrease as $\theta \rightarrow 0$ or $\pi$, where $\tau_{p}$ drops. A localized cloud of $z/z_{\rm eq} > 1$ appears slightly above the midplane but remains limited in extent. As $\theta \rightarrow 0$ or $\pi$, $z/z_{\rm eq}$ gradually falls below unity, contrasting with the behavior seen in high accretion rate models. Although disk winds can transport pairs from the midplane into the jet and upper corona, the number of pairs available for advection is low due to the smaller $\tau_{p}$ in this model. Interestingly, $z/z_{\rm eq} \sim 1$ could be observed in the jet and upper corona, which might suggest that pair processes are active even in regions of low $\tau_{p}$, restoring $z$ to its equilibrium value. However, given that the pair equilibrium timescale is significantly longer than other relevant dynamical timescales, local pair processes are expected to be negligible in these regions. Instead, the observed matching of $z$ to $z_{\rm eq}$ likely results from advection, with pairs being transported from the disk into the jet and upper corona - a coincidence rather than a signature of local equilibrium.

%==============================================================================================================%

\subsection{The Role of Advection} \label{subsec:advect}
%==============================================================================================================%

We illustrate the pair advection process by plotting time series of positron number density maps, starting from the onset of the pair injection phase. Each plot includes velocity field lines, computed following the methods described in \citet{dhang2023magnetic, chan2025impactplasmaangularmomentum}:
\begin{equation} \label{eqn:velocity}
\begin{aligned}
    V_{r} = \sqrt{g_{rr}}V^{r}, V_{\theta} &= \sqrt{g_{\theta\theta}}V^{\theta}, V_{\phi} = \sqrt{g_{\phi\phi}}V^{\phi} 
\end{aligned}
\end{equation}
where $V^{i} = u^{i}/u^{t}$ represents the $3$-velocity of the plasma. 

We begin by examining the high accretion rate model, hr01\_high, shown in Figure \ref{fig:pair_advect_tseries}(a). At $t \approx 20000$\,$GM/c^{3}$, pairs are near equilibrium in the thin strip, resulting in a high positron number density along the strip. Between $t \approx 20000$\,$GM/c^{3}$ and $20100$\,$GM/c^{3}$, turbulence that is associated with eddies in the velocity field lines begins transporting pairs into the upper corona. Then, between $t \approx 20100$\,$GM/c^{3}$ and $20400$\,$GM/c^{3}$, some pairs close enough to the black hole sneak through the jet-disk boundary (likely mediated by Kelvin-Helmholtz-like vortices at the jet-disk interface: \citealt{2024arXiv241019061V}) are launched by the bipolar outflow (shown as outward-pointing velocity field lines). A significant portion of the computational domain then becomes filled with pairs. Interestingly, some pairs are initially transported to the upper corona but are later captured by inflowing velocity fields that point toward the jet base. 

We also illustrate the advection processes for moderate accretion rate model hr02\_hot in Figure \ref{fig:pair_advect_tseries} (b). At each timestep, the positron number density reaches its spatial maximum at the midplane. However, due to its lower proton scattering optical depth, the plasma at the midplane takes longer than the high accretion rate model to attain its maximum allowable value. Still, pairs are transported by the bulk flow, filling the entire computational domain, just as in the high accretion rate model. This process likely produces the spatially limited cloud where $z/z_{\rm eq} > 1$.

The effectiveness of advection can be quantitatively addressed by comparing the inflow timescale to the pair equilibrium timescale, with relevant results shown in Figure \ref{fig:compare_tscale}. Here, the inflow timescale is computed as $t_{\rm inflow} = |r/v^{r}|$, with $v^{r} = u^{r}/u^{t}$, and we focus on the ratio $t^{r} = t_{\rm eq}/t_{\rm inflow}$. We compute $t_{\rm eq}$ using the strategy outlined in Section \ref{subsec:tscale}. We focus on model hr01\_high and consider relevant cases where the initial $z$ is both above $z_{\rm eq}$. Close to the pair-maximized thin strip $t^{r} \ll 1$, whereas in the upper corona and jet, the pair equilibrium timescale is comparatively much longer. Consequently, the pairs within the thin strip remain locked in near equilibrium, while those advected into the upper corona and jet survive without substantial annihilation, maintaining a pair fraction well above equilibrium.

A point to note: Table \ref{tab:params} shows that models with a moderate accretion rate ($H/r = 0.2$) exhibit a $z_{\rm max}$ value closer to the maximum equilibrium value, $(z_{\rm eq})_{\rm max}$, compared to those with a high accretion rate ($H/r = 0.1$). We attribute this difference to variations in flow velocity and pair generation rates between the two models. It is possible that in the high accretion rate cases, the combination of outflows and higher positron number density may lead to an enhanced positron mass flux, which in turn affects the local pair production rate within the thin strip and alters the resulting $z$ value in the dynamical model. As a result, the plasma in the thin strip settles into a new pair equilibrium state, modified from the one-zone equilibrium model by the inclusion of advection. 

%==============================================================================================================%

\subsection{The Plunging Region} \label{subsec:isco}
%==============================================================================================================%
%#####################################################################################################################%
\begin{figure}
    \centering
    \includegraphics[width=1.0\linewidth]{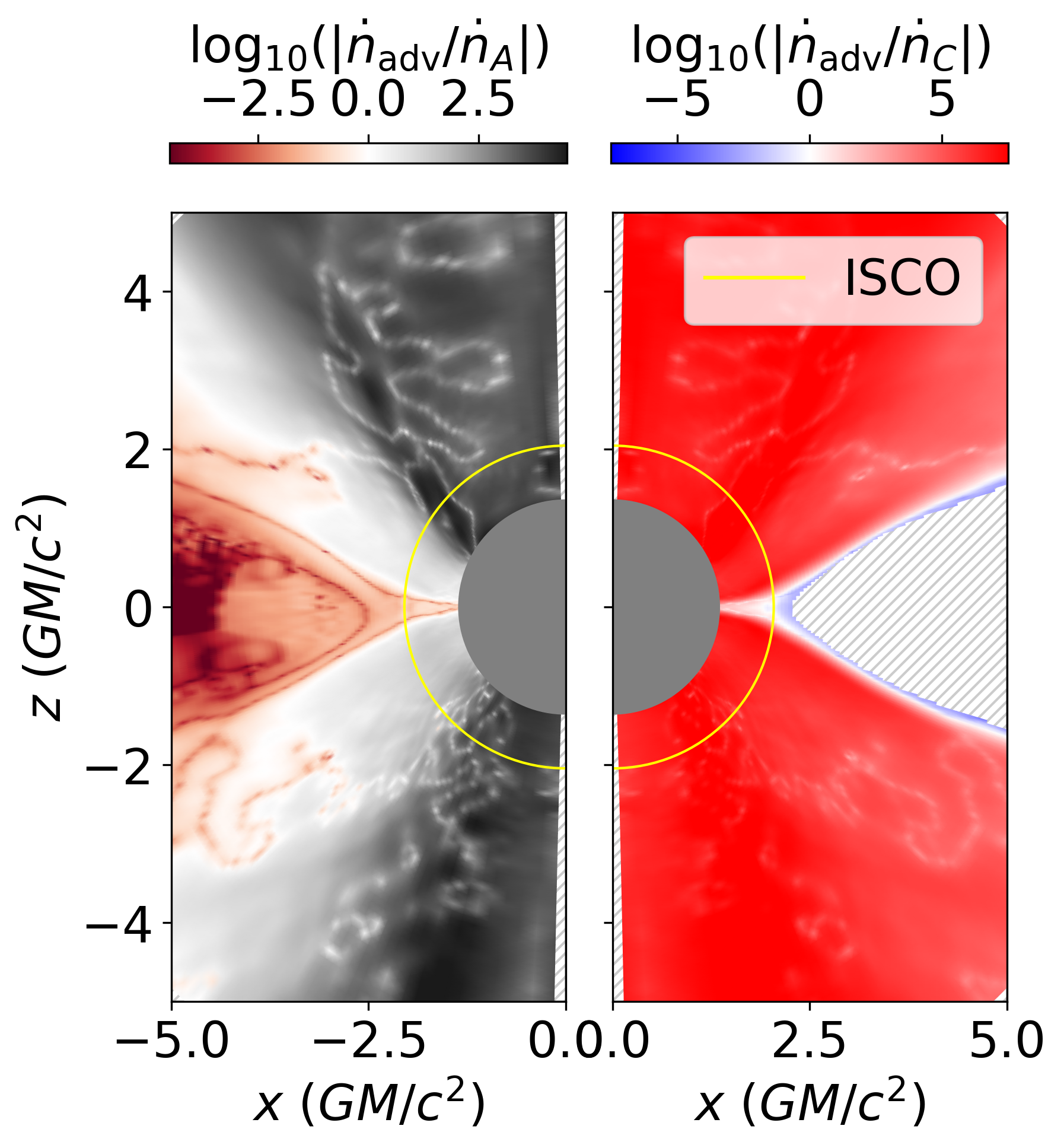}
    \caption{Ratio of the advection term to the local pair annihilation rate (left) and to the local pair creation rate (right), shown on a $\log_{10}$ scale for the high accretion rate model hr01\_high. The absolute values of the ratios are plotted, and the radius of the ISCO is indicated. Hatched regions denote the absence of a pair equilibrium solution, where the pair creation rate is not computed. \label{fig:hr01_advect}}
\end{figure}
%#####################################################################################################################%
 
We extract several radial profiles for the highest accretion rate model, hr01\_high, and present them in Figures \ref{fig:profiles_hr01}. These profiles are evaluated at selected polar angles, with our analysis restricted to the upper hemisphere for simplicity. Throughout the radial profile plots, we consistently mark the radius of the innermost stable circular orbit (ISCO; see Equation 2.21 in \citet{1972ApJ...178..347B}) with a vertical grey dotted line, a convention maintained across all relevant figures. 

We begin by examining the radial profiles of $z$. Near the midplane, the pair fraction - and correspondingly the positron number density - peaks just outside the plunging region. This occurs because $\tau_{p}$ approaches the critical threshold $\tau_{\rm crit}$, allowing local pair processes to drive $z$ to its maximum attainable value. We observe that $z$ decreases both as $r$ decreases toward the black hole and as it increases outward. The slight reduction at small $r$ is due to the drop in $\tau_{p}$, which diminishes the pair production rate. At large $r$, the decline in $z$ is attributed to $\tau_{p}$ exceeding $\tau_{\rm crit}$, thereby suppressing pair production and permitting only annihilation. Additionally, $z$ becomes smaller with increasing latitude, though its radial profile flattens. This behavior arises because $\tau_{p}$ decreases with increasing $\theta$, and remains too low to reach $\tau_{\rm crit}$. As a result, $z$ does not exhibit the sharp transitions typically associated with local pair processes.

The $z/z_{\rm eq}$ profiles exhibit distinct radial behavior. Near the midplane, $z/z_{\rm eq}$ increases as $r$ decreases toward the black hole, reaching values well above unity, but drops sharply at larger radii. In the pair runaway region, $z_{\rm eq}$ is undefined; hence, we manually set $z/z_{\rm eq} = 0$ in these zones. The resulting sharp decline in $z/z_{\rm eq}$ at large $r$ is therefore a numerical artifact rather than a physical feature. Notably, $z/z_{\rm eq} \sim 1$ just outside the plunging region - precisely where $z$ reaches its maximum across all $r$ at a given $\theta$. This behavior is, again, driven by $\tau_{p}$ approaching the critical threshold $\tau_{\rm crit}$, triggering rapid local pair processes.

Interestingly, $z$ in the plunging region is predicted to remain well above its equilibrium value. We attribute this to the influence of advection, particularly the radial infall of plasma as it crosses the ISCO. To illustrate this effect, we present the radial positron mass flux, $-\sqrt{-g}\rho_{+}u^{r}$, in Figure \ref{fig:profiles_hr01}. The flux is positive and increases at small $r$, indicating that inward transport of positrons is driven by the plunging motion. Within the plunging region, $z/z_{\rm eq}$ in the midplane is notably lower than at higher altitudes. This behavior arises because $\tau_{p}$ is of order unity there, making local pair processes non-negligible (but subdominant) relative to advection.

To qualitatively illustrate the competition between the advection term and local pair processes, we compute their respective contributions as follows:
\begin{equation} \label{eqn:pair}
\begin{aligned}
    \dot{n}_{\rm adv} &= - \partial_{r}(\sqrt{-g}\rho_{+} u^{r}) - \partial_{\theta}(\sqrt{-g}\rho_{+} u^{\theta}), \\
    \dot{n}_{\rm C} &= \sqrt{-g}\dot{\rho}_{+, C}, \\
    \dot{n}_{\rm A} &= \sqrt{-g}\dot{\rho}_{+, A}.
\end{aligned}
\end{equation}
note that we omit particle transport along the $\phi$ direction, as $\phi$-averaging has already been performed. The results are shown in Figure \ref{fig:hr01_advect} for model hr01\_high. Near the midplane, as the flow crosses the ISCO, the advection term becomes comparable to local pair processes - both creation and annihilation. Just outside the ISCO (but still not within the pair runaway region), local pair processes dominate over advection. This is because the influence of plunging plasma is less pronounced beyond the ISCO. Meanwhile, $\tau_{p}$ increases rapidly, approaching its critical threshold. As a result, the radial infall of positrons becomes comparatively less significant, while the timescale for local pair processes shortens substantially. 
%%%%%%%%%%%%%%%%%%%%%%%%%%%%%%%%%%%%%%%%%%%%%%%%%%%%%%%%%%%%%%%%%%%%%%%%%%%%%%%%%%%%%%%%%%%%%
%We note that Figure \ref{fig:hr01_advect} is consistent with Figure \ref{fig:compare_tscale} in that the advection term dominates over local pair processes in the upper corona and jet. 
%%%%%%%%%%%%%%%%%%%%%%%%%%%%%%%%%%%%%%%%%%%%%%%%%%%%%%%%%%%%%%%%%%%%%%%%%%%%%%%%%%%%%%%%%%%%%

%==============================================================================================================%

\subsection{Contribution of Pair Rates} \label{subsec:decompose}
%==============================================================================================================%
%#####################################################################################################################%
\begin{figure}
    \centering
    \includegraphics[width=1.0\linewidth]{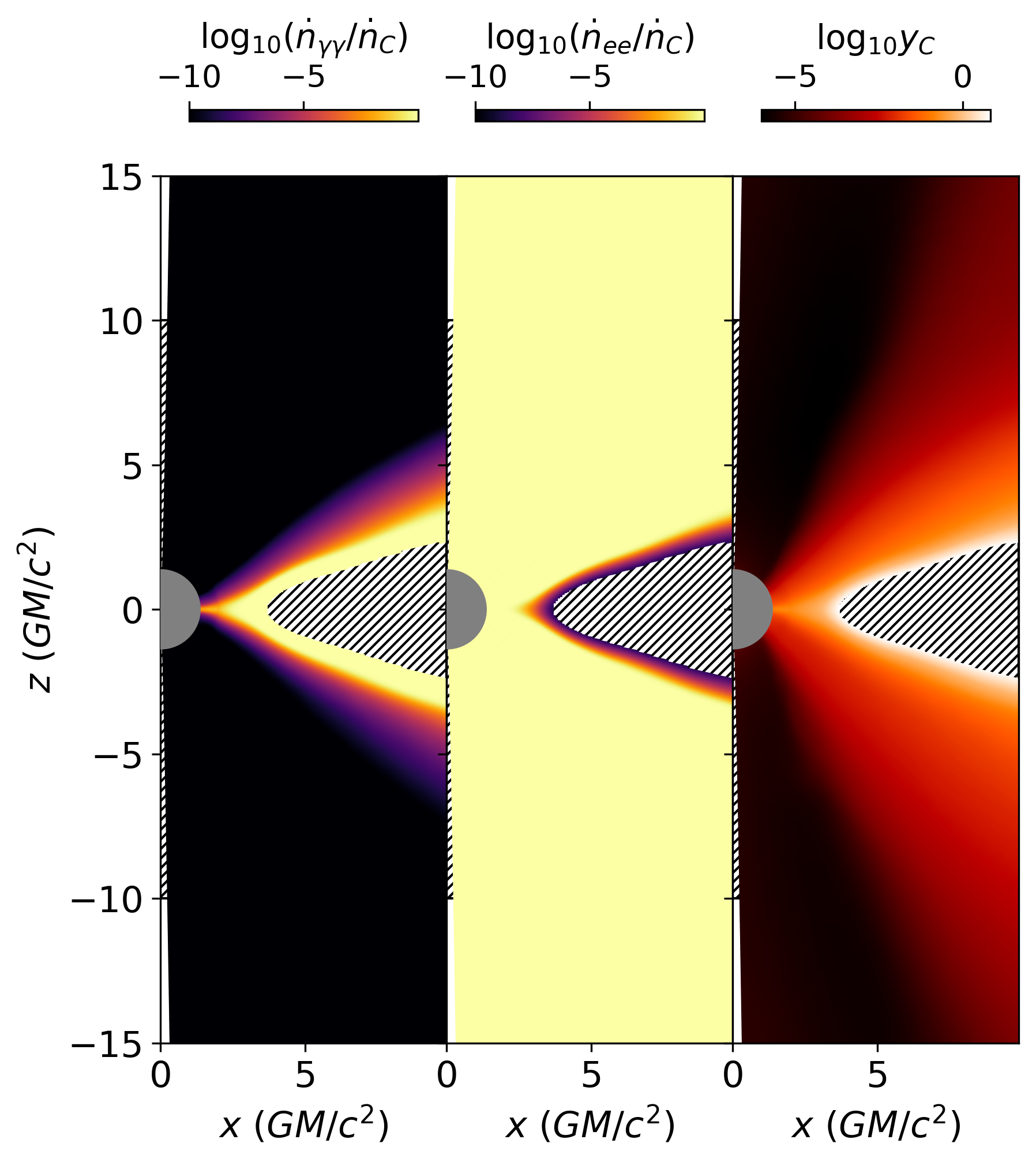}
    \caption{Ratios of photon-photon ($\dot{n}_{\gamma\gamma}$) and particle-particle ($\dot{n}_{ee}$) pair production rates relative to the total pair creation rate, shown on a $\log_{10}$ scale for the high accretion rate model hr01. Hatched regions indicate the absence of a pair equilibrium solution. Additionally, the non-relativistic Compton parameter is displayed on a $\log_{10}$ scale using a distinct colormap. \label{fig:pair_decompose_ih3d_hr01_pp_reduced}}
\end{figure}
%#####################################################################################################################%

The pair production rates contain contributions from photon-photon ($\dot{n}_{\gamma\gamma}$), photon-particle ($\dot{n}_{\gamma e}$), and particle-particle ($\dot{n}_{ee}$) processes. Here, we examine which pair production process dominates in different spatial regions of the accretion flow. To achieve this, we decompose the pair production rates into their respective contributions and present the results in Figure \ref{fig:pair_decompose_ih3d_hr01_pp_reduced} (a), focusing on the high accretion rate model hr01. Close to the midplane but outside the pair runaway region (which is shown as a hatched region), the photon-photon process is the dominant source of pair production. In contrast, in the upper corona and jet, the particle-particle process dominates. The photon-particle process is significant in a limited spatial region but remains subdominant compared to the photon-photon process, so we omit it from the figure. Additionally, in the same figure, we include a plot of the Compton parameter:
\begin{equation}
    y_{C} = \frac{4k_{B}T_{e}}{m_{e}c^{2}}\text{min}(\tau_{T}, \tau_{T}^{2}),
\end{equation}
with $k_{B}$ being the Boltzmann constant. The Compton parameter increases toward the midplane, where the photon-photon process peaks. There, Compton scattering becomes an important mechanism for energizing photons. This suggests that, near the midplane, photons participating in pair-producing collisions are most likely upscattered to the Wien peak. We also note that the total pair production rate peaks near the midplane. Thus, even though the particle-particle process dominates in the upper corona and jet, the absolute value of the corresponding rate remains low in these regions. The dominance of the photon-photon interaction rate declines sharply toward the pole. This rapid drop arises because the rate depends on the number of photons in the Wien peak, $n_{\gamma}$, which itself depends on the fraction of upscattered photons, $f_{B}$. The latter scales approximately as $\tau_{T}^{4}$, neglecting the exponential suppression term, which also tends toward zero as $\tau_{T} \rightarrow 0$. Lastly, the moderate accretion rate model hr02\_hot is similar to model hr01, in that we can still observe regions where photon-photon rates exist and remain dominant. However, in the low accretion rate model, i.e., adaf, the pair production process is completely dominated by particle-particle interactions.

%==============================================================================================================%

\subsection{Goldreich-Julian Charge Density} \label{subsec:ngj}
%==============================================================================================================%
%#####################################################################################################################%
\begin{figure}
    \centering
    \includegraphics[width=1.0\linewidth]{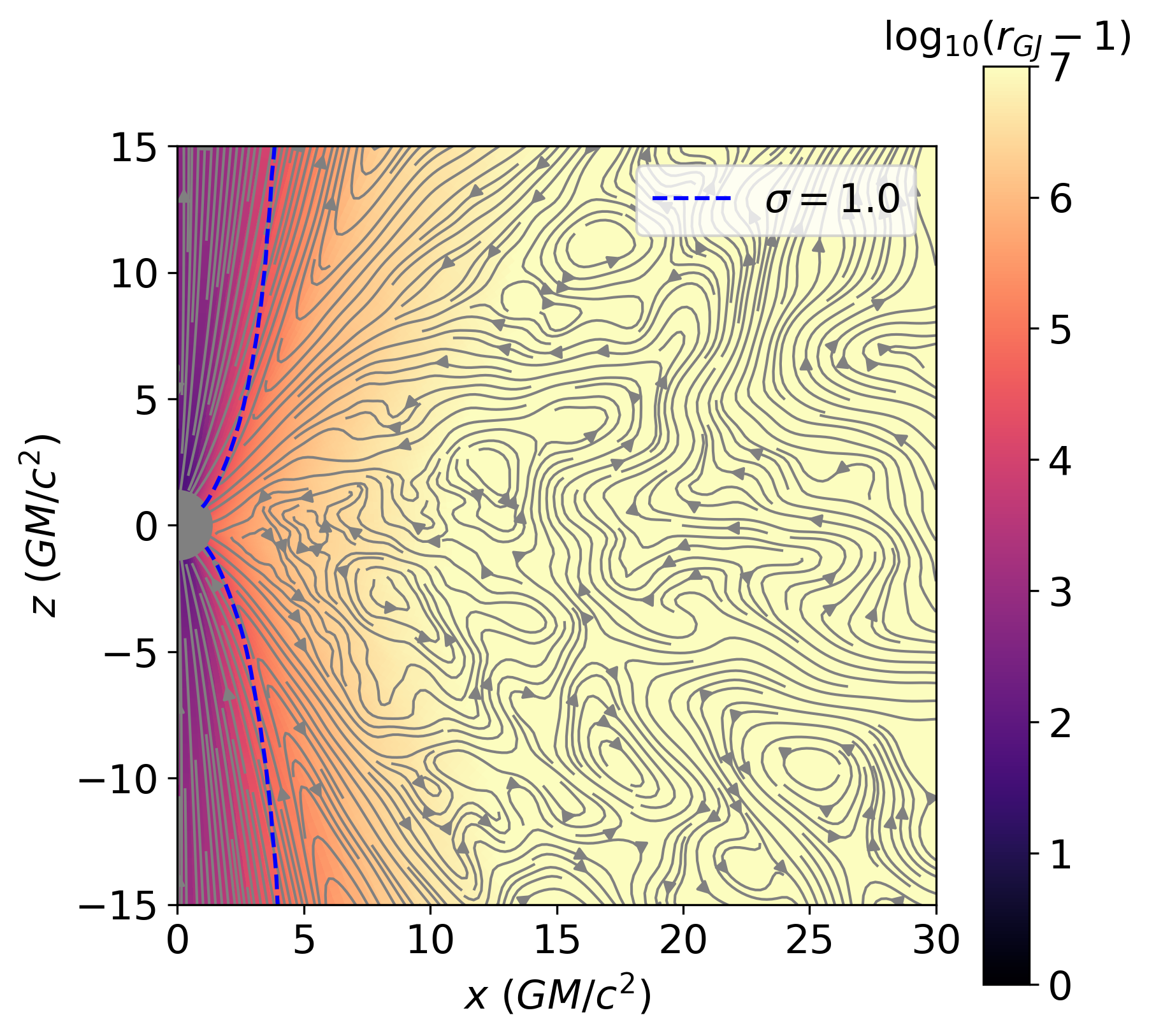}
    \caption{The ratio of positron number density to the Goldreich–Julian density, $r_{\rm GJ} = n_{+}/n_{\rm GJ}$, is shown for the low accretion rate model adaf. To better highlight regions where $r_{\rm GJ} < 1$, we plot $r_{\rm GJ} - 1$ on a $\log_{10}$ scale. No undefined regions are found, confirming that $r_{\rm GJ} > 1$ throughout the domain. For reference, we include the contour where $\sigma = 1$ and overlay magnetic field lines. \label{fig:ngj_model}}
\end{figure}
%#####################################################################################################################%
 
We examine whether pair advection from the disk can supply the bipolar region with sufficient charge density to screen the electric field. The necessary density, known as the Goldreich–Julian density \citep{1969ApJ...157..869G, 2011ApJ...735....9M, 2021ApJ...907...73W, 2021MNRAS.507.4864Y}, is given by:
\begin{equation} 
\begin{aligned}
    n_{GJ} = -u^{\mu}j_{\mu}, \\
    j^{\mu} = \nabla_{\nu}F^{\mu\nu}.
\end{aligned}
\end{equation}
where $F^{\mu\nu}$ denotes the Faraday tensor and $\nabla_{\nu}$ the covariant derivative. The computation of the four-current $j^{\mu}$ requires a time derivative; in \texttt{iharm3d}, this is performed at each timestep by finite differencing $F^{\mu\nu}$ in the spatial-temporal coordinates. We compare the pair number density to the Goldreich–Julian density and present the results in Figure \ref{fig:ngj_model} for the low accretion rate model adaf. For completeness, we also include magnetic field lines, computed following \citet{dhang2023magnetic} and \citet{chan2025impactplasmaangularmomentum}:
\begin{equation} \label{eqn:bfield}
\begin{aligned}
    B_{r} = \sqrt{g_{rr}}B^{r}, B_{\theta} &= \sqrt{g_{\theta\theta}}B^{\theta}, B_{\phi} = \sqrt{g_{\phi\phi}}B^{\phi} 
\end{aligned}
\end{equation}
and that we compute the ratio of positron number density to the Goldreich–Julian density as $r_{\rm GJ} = n_{+}/n_{\rm GJ}$. We find that $r_{\rm GJ} > 1$ throughout the domain. In particular, this ratio exceeds unity in the strongly magnetized ($\sigma > 1$) bipolar region, which is concentrated with poloidal magnetic fields. However, the ratio is comparatively lower than that in the disk, because bipolar jets carry stronger currents than the disk. 

%==============================================================================================================%

\subsection{Coulomb Collision Timescale} \label{subsec:coulumb}
%==============================================================================================================%
%#####################################################################################################################%
\begin{figure}
    \centering
    \includegraphics[width=1.0\linewidth]{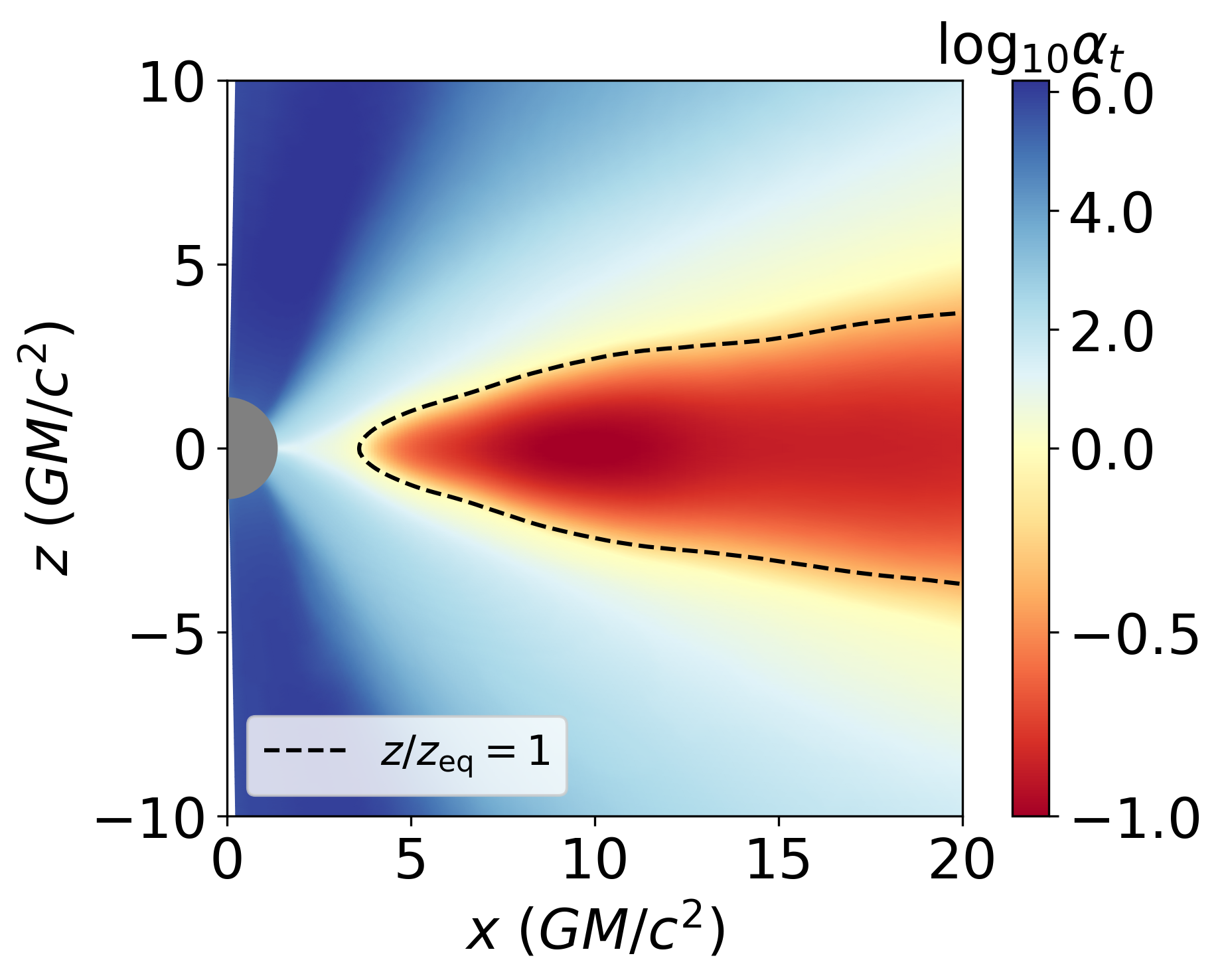}
    \caption{The Coulomb efficiency parameter, $\alpha_{t}$, is shown on a $\log_{10}$ scale for the high accretion rate model hr01, which features a high accretion rate. Contours corresponding to $z/z_{\rm eq} = 1$ are overlaid on the plot. The $z/z_{\rm eq} = 1$ contour encloses the region where $\alpha_{t} < 1$, indicating that the pair void coincides with this low Coulomb energy exchange efficiency zone. \label{fig:columb_time}}
\end{figure}
%#####################################################################################################################%
 
We verify that our decision to manually disable pair production based on $\tau_{\rm crit}$ - at least for the model presented in this study - is self-consistent. To do so, we assess the efficiency of Coulomb energy transfer to determine whether electrons are sufficiently cooled, thereby suppressing pair production. This is achieved by comparing the Coulomb collision timescale, $t_{\rm coul}$, to the orbital timescale, $t_{\Omega} = 1/\Omega$. We then compute the ratio $\alpha_{t} = t_{\rm coul}/t_{\Omega}$, which indicates whether Coulomb energy exchange can occur within a single orbital period. The Coulomb collision timescale, $t_{\rm coul}$, is given by:
\begin{equation} 
    t_{\rm coul} = \text{min}(\frac{u_{\pm}}{\dot{U}_{\rm coul}}, \frac{u_{p}}{\dot{U}_{\rm coul}}),
\end{equation}
in which $\dot{U}_{\rm coul}$ is the Coulomb energy exchange rate \citep{stepney1983numerical}:
\begin{equation} \label{eqn:columb}
\begin{aligned}
    & \dot{U}_{\rm coul} = -\frac{3m_{e}}{2m_{p}}(n_{e} + n_{+})n_{p}\sigma_{t}c\frac{k_{B}(T_{e} - T_{p})}{K_{2}(1/\theta_{e})K_{2}(1/\theta_{p})}\text{ln}\Lambda \\
    & \times \left[\frac{2(\theta_{e}+\theta_{p})^{2}+1}{\theta_{e}+\theta_{p}}K_{1}\left(\frac{\theta_{e}+\theta_{p}}{\theta_{e}\theta_{p}}\right) + 2K_{0}\left(\frac{\theta_{e}+\theta_{p}}{\theta_{e}\theta_{p}}\right) \right],
\end{aligned}
\end{equation}
Here, $\theta_{e,p} = k_{B}T_{e,p}/m_{e,p}c^{2}$ denotes the dimensionless temperature for electrons and protons, respectively. The internal energy per unit volume is given by $u_{g} = u_{p} + u_{e}$, where $u_{p} = P/(\gamma - 1)$ represents the proton contribution, and $u_{\pm} = 3(n_{e} + n_{+})k_{B}T_{e}$ accounts for electrons and positrons. The Coulomb logarithm is approximated as $\Lambda \approx 20$, and $K_{n}$ denotes the modified Bessel function of the second kind of order $n$. We present the Coulomb efficiency parameter, $\alpha_{t}$, in Figure \ref{fig:columb_time}, highlighting results from the high accretion rate model hr01. We find that $\alpha_{t}$ falls below unity, particularly near the midplane. The $z/z_{\rm eq} = 1$ contour encloses the region where $\alpha_{t} < 1$ and approximately coincides with the $\alpha_{t} = 1$ boundary. Notably, this contour lies just above the pair void - the region where pair runaway is expected to occur and where pair production is manually disabled. In contrast, the lower accretion rate model hr02\_hot, which lacks a pair void, consistently exhibits $\alpha_{t} > 1$. In fact, the minimum value of $\alpha_{t}$ in this model is on the order of $\mathcal{O}(10)$. 

%==============================================================================================================%

\subsection{The Impact of Mass Accretion Rate} \label{subsec:rateimpact}
%==============================================================================================================%
%#####################################################################################################################%
\begin{figure}
    \centering
    \includegraphics[width=1.0\linewidth]{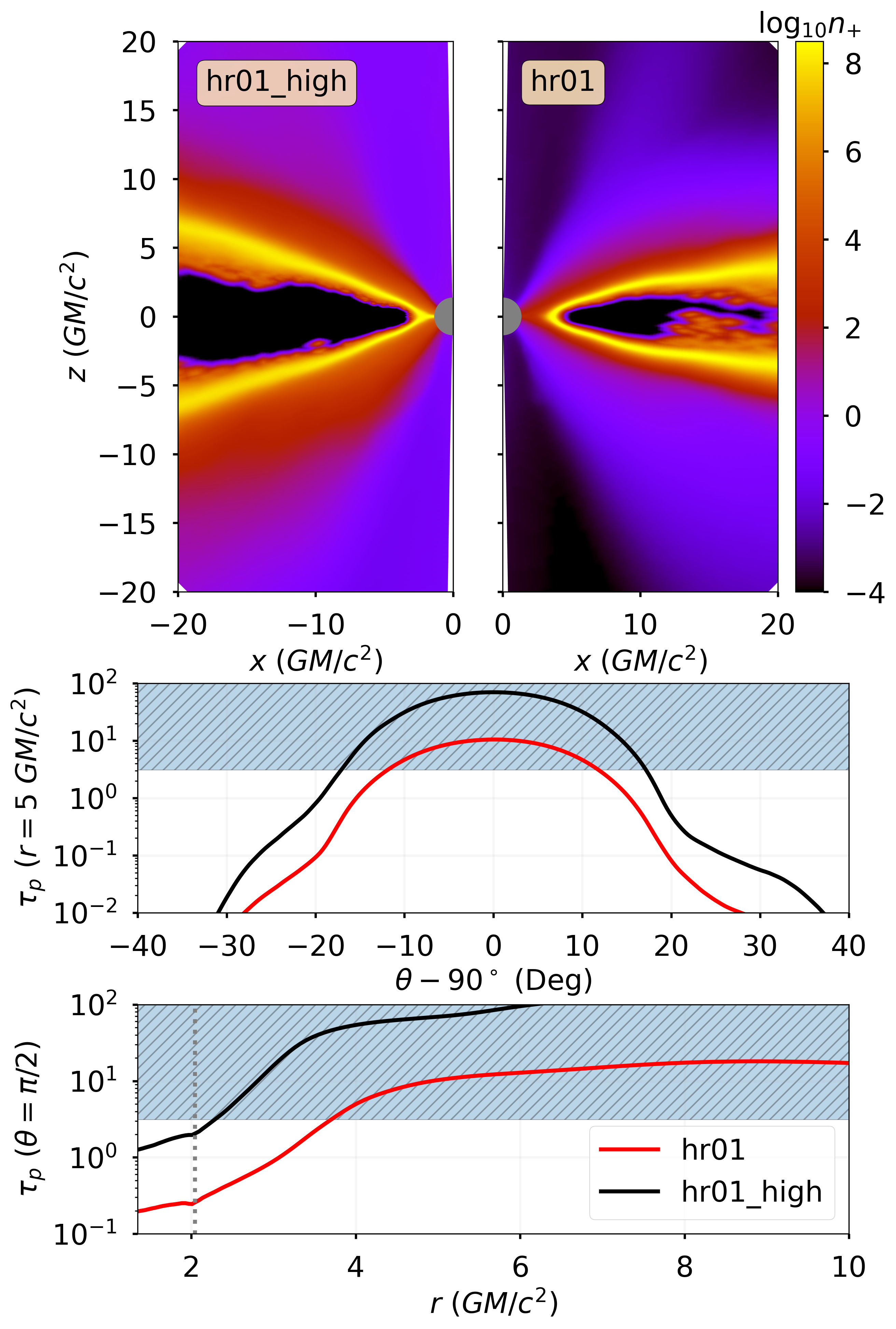}
    \caption{Upper panel: Spatial distribution of positron number density for model hr01\_high (left, $\eta = 0.4$) and model hr01 (right, $\eta = 0.05$), shown using a shared colormap. Middle panel: Comparison of $\tau_{p}$ as a function of $\theta$ at $r = 5$\,$GM/c^{2}$ for models hr01 and hr01\_high. The shaded region indicates where $\tau_{p} \geq \tau_{\rm crit}$. Bottom panel: Same as the middle panel, but showing the midplane $\tau_{p}$. The vertical grey line marks the ISCO. The middle and bottom panels share the same legend. \label{fig:mdotimpact}}
\end{figure}
%#####################################################################################################################%

We introduce an additional model, hr01\_high, which features a mass accretion rate eight times greater than that of hr01. Despite this elevated accretion rate, the maximum pair fraction remains modest at $z_{\rm max} = 4.753\times10^{-3}$, and the positron number density is not significantly enhanced (see Table \ref{tab:params}), as the upper limit is constrained by the cutoff $\tau_{\rm crit}$. Figure \ref{fig:mdotimpact} compares hr01\_high and hr01. In the high accretion model, the pair-maximized thin strip shifts closer to the black hole horizon, exhibits a larger opening angle, and contains a greater number of pairs within the computational domain. This behavior arises because the proton scattering optical depth scales with accretion rate. As a result, the location where $\tau_{p}$ reaches $\tau_{\rm crit}$ moves inward and upward, repositioning the pair-maximized strip to smaller radii and expanding its angular extent. Under these conditions, pairs are more readily captured by strong bipolar outflows and disk winds, promoting more efficient advection into the upper corona and jets. We also note that $z_{\rm max}$ is slightly lower in the model with a higher accretion rate, likely due to the influence of outflows.  

%==============================================================================================================%

\subsection{The Impact of Black Hole Mass} \label{subsec:bhmass}
%==============================================================================================================%
%#####################################################################################################################%
\begin{figure}
    \centering
    \includegraphics[width=1.0\linewidth]{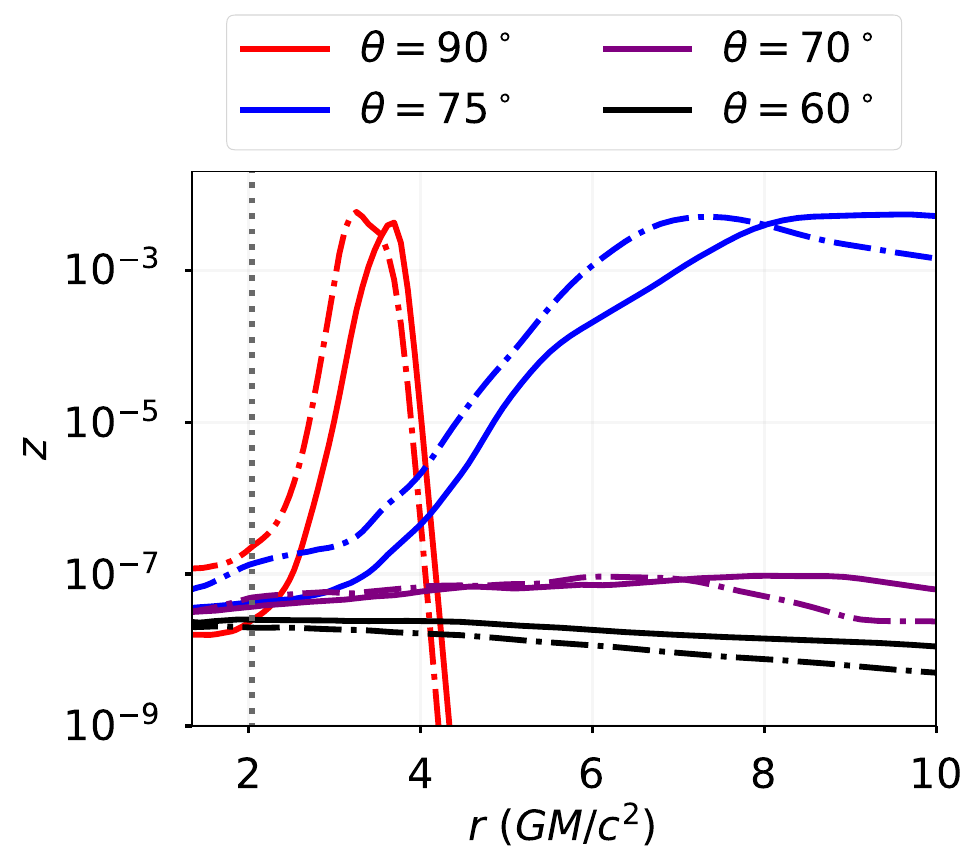}
    \caption{Radial profiles of $z = n_{+}/n_{p}$, sampled at various $\theta$ angles in the upper hemisphere, are shown for the $M_{\rm BH} = 10^{8}$\,$M_{\odot}$ model hr01 (solid lines) and the $M_{\rm BH} = 10$\,$M_{\odot}$ model hr01\_10m (dash-dotted lines). The vertical dotted grey line indicates the location of the ISCO. \label{fig:compare_10m_model}}
\end{figure}
%#####################################################################################################################%

We investigate how the results depend on black hole mass by introducing an additional model, hr01\_10m, which is identical to hr01 except that the black hole mass is reduced to $10$\,$M_{\odot}$. As shown in Table \ref{tab:params}, both models yield comparable values for $(z_{\rm eq})_{\rm max}$, $z_{\rm max}$, and $(z/z_{\rm eq})_{\rm max}$. The model with the larger black hole mass exhibits a higher positron number density, primarily because the spatial scale - characterized by its typical length $GM_{\rm BH}/c^{2}$ - is smaller. We also compare the radial profiles of $z$ for both models in Figure \ref{fig:compare_10m_model}, finding them to be quite similar. The observed differences can be attributed to the weak dependence of the one-zone pair production model on source size. An analytic argument supporting this is provided in Appendix \ref{subapp:dependence}. Importantly, these minor differences in the one-zone model do not amplify in global simulations, even when the black hole mass spans seven orders of magnitude. 

%==============================================================================================================%

\section{Implications} \label{sec:imply}

%==============================================================================================================%

\subsection{The Pair Timescale Governs the Spatial-Temporal Distribution of Pairs} \label{subsec:advecdominate}
%^^^^^^^^^^^^^^^^^^^^^^^^^^^^^^^^^^^^^^^^^^^^^^^^^^^^^^^^^^^^^^^^^^^^^^^^^^^^^^^^^^^^^^^^^^^^^^^^^^^^^^^^^^^^^^%
Our results underscore the physics of pair production as a competition between the timescales of local pair processes and global dynamics. The one-zone calculations support our treatment of electron-positron pairs as an advected passive scalar in global simulations. In regions of low proton scattering optical depth, pair processes are likely influenced by global accretion dynamics, whereas in regions of high optical depth, pairs tend to be locked into local pair equilibrium. The long pair equilibrium timescale at small $\tau_{p}$ arises from the dominance of thermal particle-particle interactions, which are inefficient at producing pairs due to their small collision cross-section.

%%%%%%%%%%%%%%%%%%%%%%%%%%%%%%%%%%%%%%%%%%%%%%%%%%%%%%%%%%%%%%%%%%%%%%%%%%%%%%%%%%%%%%%%%%%%%%%%%%%%%%%%%%%%%%%%%%%%%%%%%%%%%%%%%%%%%%%%%%%%%%%%%%%%%%%%%%%%%%%
%Note that a purely thermal pair plasma does not generate an annihilation line feature in the emergent spectrum \citep{1986ApJ...303...94Z}, because its spectrum is either intrinsically broad or strongly smeared by repeated Compton scatterings \citep{1995MNRAS.276..273M}. Still, we do not expect photons injected via pair annihilation to be the dominant emission mechanism, since observations indicate that the spectra of accreting black holes \citep{2006ARA&A..44...49R, 2007A&ARv..15....1D} peak at $1$\,keV in the soft state or exhibit an exponential cutoff at $100$\,keV in the hard state. Therefore, our model should be consistent with observational constraints.
%%%%%%%%%%%%%%%%%%%%%%%%%%%%%%%%%%%%%%%%%%%%%%%%%%%%%%%%%%%%%%%%%%%%%%%%%%%%%%%%%%%%%%%%%%%%%%%%%%%%%%%%%%%%%%%%%%%%%%%%%%%%%%%%%%%%%%%%%%%%%%%%%%%%%%%%%%%%%%%

The hierarchy of pair-equilibrium timescales governs the spatial and temporal distribution of pairs in our model. In high accretion rate cases, for example, continuous transport of pairs from the pair-maximized thin strip persists due to the high value of $\tau_{p}$, which enables rapid pair regeneration even as pairs are depleted by particle transport. The thin strip thus acts as a persistent pair source in the model; Within the pair runaway region, pair annihilation dominates over advection by several orders of magnitude, leading to the prompt destruction of any pairs advected into this region. Consequently, a pair void emerges rather than a smooth decline in pair density toward the midplane; The dynamics of pairs in the plunging region are governed by the strong advection term arising from the plunging flow. Thus, local pair regulation becomes less effective - even when $\tau_{p}$ is of order unity. 

%%%%%%%%%%%%%%%%%%%%%%%%%%%%%%%%%%%%%%%%%%%%%%%%%%%%%%%%%%%%%%%%%%%%%%%%%%%%%%%%%%%%%%%%%%%%%%%%%%%%%%%%%%%%%%%%%%%%%%%%%%%%%%%%%%%%%%%%%%%%%%%%%%%%%%%%%%%%%%%
%The physics of the plunging region around black holes has recently attracted significant attention, prompting studies that predict continuum emission \citep{wilkins2020venturing, mummery2024continuum}, power-law spectra \citep{hankla2022non}, and polarization signatures \citep{chan2025polarimetricsignaturesbulkcomptonization} originating from this zone. 
%%%%%%%%%%%%%%%%%%%%%%%%%%%%%%%%%%%%%%%%%%%%%%%%%%%%%%%%%%%%%%%%%%%%%%%%%%%%%%%%%%%%%%%%%%%%%%%%%%%%%%%%%%%%%%%%%%%%%%%%%%%%%%%%%%%%%%%%%%%%%%%%%%%%%%%%%%%%%%%

\subsection{The Impact of Model Parameters} \label{subsec:impactmodel}
%^^^^^^^^^^^^^^^^^^^^^^^^^^^^^^^^^^^^^^^^^^^^^^^^^^^^^^^^^^^^^^^^^^^^^^^^^^^^^^^^^^^^^^^^^^^^^^^^^^^^^^^^^^^^^^%
By varying model parameters, we demonstrate their impact on the results - the effectiveness of pair advection processes appears to depend on the system’s mass accretion rate. Conversely, our findings are not highly sensitive to the assumed black hole mass, suggesting that the results should be equally applicable to accretion systems in both the AGN and X-ray binary regimes, which also feature coronae \citep{1996ApJ...470..249P}.

%%%%%%%%%%%%%%%%%%%%%%%%%%%%%%%%%%%%%%%%%%%%%%%%%%%%%%%%%%%%%%%%%%%%%%%%%%%%%%%%%%%%%%%%%%%%%%%%%%%%%%%%%%%%%%%%%%%%%%%%%%%%%%%%%%%%%%%%%%%%%%%%%%%%%%
%Comparing the observed compactness-temperature distribution of AGN with the pair equilibrium line shows that most AGN respect the pair equilibrium constraint \citep{2015MNRAS.451.4375F,2017MNRAS.467.2566F,2019BAAS...51c.126K}. Additionally, requiring the accreting plasma around X-ray binaries to satisfy the pair equilibrium condition yields a corresponding radius of pair dominance ($z = 1$) of approximately $4$\,$GM/c^{2}$ \citep{zdziarski2021hybrid}, which is close to the characteristic length scale of a black hole. 
%%%%%%%%%%%%%%%%%%%%%%%%%%%%%%%%%%%%%%%%%%%%%%%%%%%%%%%%%%%%%%%%%%%%%%%%%%%%%%%%%%%%%%%%%%%%%%%%%%%%%%%%%%%%%%%%%%%%%%%%%%%%%%%%%%%%%%%%%%%%%%%%%%%%%%

\subsection{Pair Processes may Regulate Coronal Temperature} \label{subsec:pairregulate}
%^^^^^^^^^^^^^^^^^^^^^^^^^^^^^^^^^^^^^^^^^^^^^^^^^^^^^^^^^^^^^^^^^^^^^^^^^^^^^^^^^^^^^^^^^^^^^^^^^^^^^^^^^^^^^^%

Empirical findings raise questions about whether pair physics plays a significant role in regulating plasma temperature \citep{2015MNRAS.451.4375F,zdziarski2021hybrid}. Our current model takes a first step in addressing this question theoretically by including sub-grid prescriptions for positron evolution in GRMHD. However, it cannot provide a fully consistent answer, primarily because we fix the electron temperature at $T_e = 10^9$ K and neglect regions where pair-production would runaway (and cool the electrons). In principle, this could be done in full radiation GRMHD, but the optically thick disk of $T \sim 10^7$ K corresponds to a gas pressure scale height of $H/R \sim 10^{-3}$ which is currently not possible to resolve computationally.

%%%%%%%%%%%%%%%%%%%%%%%%%%%%%%%%%%%%%%%%%%%%%%%%%%%%%%%%%%%%%%%%%%%%%%%%%%%%%%%%%%%%%%%%%%%%%%%%%%%%%%%%%%%%%%%%%%%%%%%%%%%%%%%%%%%%%%%%%%%%%%%%%%%%%%
%This is due to limited computational resources, which prevent us from resolving a thin accretion disk with a temperature of $10^{7}$\,$\mathrm{K}$. A slightly better solution to this would be to assign the electron temperature based on empirical functions \citep{2024ApJ...964...17C, 2024arXiv240804132C}.
%%%%%%%%%%%%%%%%%%%%%%%%%%%%%%%%%%%%%%%%%%%%%%%%%%%%%%%%%%%%%%%%%%%%%%%%%%%%%%%%%%%%%%%%%%%%%%%%%%%%%%%%%%%%%%%%%%%%%%%%%%%%%%%%%%%%%%%%%%%%%%%%%%%%%%

Nevertheless, by examining the pair equilibrium timescale, we may arrive at a qualitative estimate. For the pair physics to serve as a temperature-regulating mechanism, its timescale must be comparable to that of the primary heating source, which in this case is the Coulomb energy exchange. Consider models with $H/r = 0.1$ as examples. The Coulomb collision timescale is comparable to the orbital timescale in the pair-maximized thin strip. The orbital time is at least $t_{\omega} \sim 2$\,$GM/c^{3}$ (assuming $r = 1$\,$GM/c^{2}$), but the pair equilibrium timescale is of a similar order of magnitude (see Figure \ref{fig:pair_onezone_svensson} (b)). Thus, the influence of pair physics cannot be ignored - it likely plays an important, if not dominant, role in regulating plasma temperature. A definitive conclusion, however, requires explicit tracking of individual species’ entropy and the inclusion of relevant cooling mechanisms. 

\subsubsection{Manual Suppression of Pair Production is Self-consistent} \label{subsubsec:verifyrunaway}
%-------------------------------------------------------------------------------------------------%
We verified that our practical approach of manually turning off pair production in the runaway region is self-consistent. In the high accretion rate model, regions of efficient Coulomb collisions coincide with the pair void associated with the runaway zone, making it reasonable to assume that both protons and electrons (including positrons) maintain a low but unified temperature of $T_{e} = T_{p} \sim 10^{7}$\,K \citep{dexter2021radiation, 2024A&A...692A.153S}. At these temperatures, the pair annihilation rate consistently exceeds the production rate, implying that any excess pairs advected into the runaway region are promptly annihilated, rendering it effectively pair-free. In contrast, the moderate accretion rate model features $\tau_{p}$ values that remain below the critical threshold, and thus we do not observe regions of efficient Coulomb energy exchange

%%%%%%%%%%%%%%%%%%%%%%%%%%%%%%%%%%%%%%%%%%%%%%%%%%%%%%%%%%%%%%%%%%%%%%%%%%%%%%%%%%%%%%%%%%%%%%%%%%%%%%%%%%%%%%%%%%%%%%%%%%%%%%%%%%%%%%%%%%%%%%%%%%%%%%
%\footnote{As such, we reframe the question from `How quickly does pair production cool the overheated corona?' to `If a pair plasma is at temperature $T_e$, how long does it take to reach equilibrium?'} 
%%%%%%%%%%%%%%%%%%%%%%%%%%%%%%%%%%%%%%%%%%%%%%%%%%%%%%%%%%%%%%%%%%%%%%%%%%%%%%%%%%%%%%%%%%%%%%%%%%%%%%%%%%%%%%%%%%%%%%%%%%%%%%%%%%%%%%%%%%%%%%%%%%%%%%

\subsection{Pair Outflows and Jets} \label{subsec:pairoutflowjet}
%^^^^^^^^^^^^^^^^^^^^^^^^^^^^^^^^^^^^^^^^^^^^^^^^^^^^^^^^^^^^^^^^^^^^^^^^^^^^^^^^^^^^^^^^^^^^^^^^^^^^^^^^^^^^^^%

\subsubsection{Self-consistent Production of Pair Outflows} \label{subsubsec:pairoutflow}
%-------------------------------------------------------------------------------------------------%
Pioneering works on the one-zone pair equilibrium model \citep{1981ApJ...251..713L, 1982ApJ...258..335S, 1982ApJ...253..842L, 1984MNRAS.209..175S} initially neglected the possibility of pair escape. However, it was later shown that allowing pair escape could affect the existence of an equilibrium disk solution \citep{1995ApJ...449..813M}. Our models demonstrate that pairs can be advected and subsequently loaded into disk outflows or jets. Given a sufficiently strong outflow, pairs may escape to large distances. Thus, black hole candidates that were initially considered too luminous to achieve an equilibrium disk solution, when accounting for pair physics, could be significantly influenced by pair escape. Additionally, this process could potentially account for the origin of positrons in certain X-ray binaries, such as 1E 1740.7–2942 \citep{1991ApJ...383L..45B, 1991ApJ...383L..49S, 1992Natur.358..215M, 2009ApJ...693.1871B}.  

%%%%%%%%%%%%%%%%%%%%%%%%%%%%%%%%%%%%%%%%%%%%%%%%%%%%%%%%%%%%%%%%%%%%%%%%%%%%%%%%%%%%%%%%%%%%%%%%%%%%%%%%%%%%%%%%%%%%%%%%%%%%%%%%%%%%%%%%%%%%%%%%%%%%%%
%Specifically, it has been suggested that electron-positron pairs are ejected from the black hole at relativistic velocities and eventually slow down and annihilate in the surrounding cold, dense molecular cloud \citep{2001ApJ...548..394C}. In our model, the survival of pairs at large distances from the black hole could be facilitated by the long annihilation timescale in a low $\tau_{p}$ medium. 
%%%%%%%%%%%%%%%%%%%%%%%%%%%%%%%%%%%%%%%%%%%%%%%%%%%%%%%%%%%%%%%%%%%%%%%%%%%%%%%%%%%%%%%%%%%%%%%%%%%%%%%%%%%%%%%%%%%%%%%%%%%%%%%%%%%%%%%%%%%%%%%%%%%%%%

\subsubsection{Screening the Electric Field Along the Jet} \label{subsubsec:screenefield}
%-------------------------------------------------------------------------------------------------%
In GRMHD models, it remains uncertain whether the particle density along bipolar jets exceeds $n_{GJ}$, the threshold required to screen the electric field. This uncertainty arises because, in strongly magnetized regions, the plasma is dominated by artificially injected mass-energy floors that ensure stable time evolution of the simulation. Unless a force-free model is employed \citep{2024MNRAS.532.3198C}, the particle number density in these regions tends to be overestimated. Our results demonstrate that advection can load the strongly magnetized bipolar region with positrons, leading to a pair density that exceeds $n_{GJ}$. We therefore propose that pair mass loading via advection could serve as a mechanism for injecting sufficient charge density to screen the jet's electric field - particularly in low-luminosity accreting black holes. We note that bipolar jets in the SANE model are weak, and it remains uncertain whether a similar mechanism can operate in the strongly magnetized MAD regime. This question is left for future investigation.

\subsubsection{Jet Composition} \label{subsubsec:composition}
%-------------------------------------------------------------------------------------------------%
Still, whether black hole jets consist primarily of pure electron-proton plasma \citep{1996MNRAS.283..873R}, pure electron-positron plasma \citep{1998Natur.395..457W, 2000ApJ...534..109S, 2016MNRAS.457.1124K, 2022ApJ...925..189Z}, or a mixture of both remains an unresolved issue \citep{2012MNRAS.424L..26G}. Observations of circular polarization in M87, for example, have been shown to be consistent with a pair fraction of $z \sim 0.11$ \citep{2021ApJ...923..272E}. Even if our model predicted pair injection along bipolar jets via advection, the pair fraction along the jets is subject to uncertainty due to imposed floor densities. For model adaf, we find $z \sim 10^{-5}$ along the jet. Thus, unless the actual proton number density is at least $10^{-4}$ lower than the value implied in GRMHD models, it is unlikely that pair injection via particle transport from the disk alone could account for a pair-dominated jet. Simulations using a hybrid force-free model \citep{2024MNRAS.532.3198C} might be able to obtain a more accurate estimate of the pair fraction. 

\subsection{Connecting to Observational Results} \label{subsec:compareobserve}
%^^^^^^^^^^^^^^^^^^^^^^^^^^^^^^^^^^^^^^^^^^^^^^^^^^^^^^^^^^^^^^^^^^^^^^^^^^^^^^^^^^^^^^^^^^^^^^^^^^^^^^^^^^^^^^%

It was suggested that accreting black hole are unlikely to be pair-dominated \citep{1997ApJ...476..620H, 1997ASPC..121..101M, 2017ApJ...847...96L, 2017MNRAS.468.3489K}. In this work, the maximum pair fraction found in models with $H/r = 0.1$ is $\sim \mathcal{O}(0.01)$, consistent with some of the recent estimations done on X-ray binaries \citep{2021ApJ...914L...5Z}. This consistency is very likely due to our implicit exclusion of the high-$z$ branch of the pair equilibrium solution. Nonetheless, the maximum pair fraction for the model with $H/r = 0.1$ occurs in $\tau_{p} \sim \mathcal{O}(1)$ regions where the plasma is locked into near pair equilibrium. Also, the high-$z$ equilibrium solution is unstable to perturbations, so any initially high-$z$ plasma would rapidly settle to the low-$z$ solution, which has an equilibrium value of $\mathcal{O}(0.01)$, unless the electrons are continuously heated. It should be stressed that during pair runaway, the pair fraction could rise to $\mathcal{O}(10^{2})$, even if the initial and final values of $z$ remain low. However, this increase would occur only over a short period of time \citep{1986Natur.319..649M, white1990instabilities}. 

\subsection{Comparison to Previous Work} \label{subsec:compareprevious}
%^^^^^^^^^^^^^^^^^^^^^^^^^^^^^^^^^^^^^^^^^^^^^^^^^^^^^^^^^^^^^^^^^^^^^^^^^^^^^^^^^^^^^^^^^^^^^^^^^^^^^^^^^^^^^^%

Previously, \citet{2011ApJ...735....9M, 2021ApJ...907...73W} studied the impact of pair physics on SANE accretion disk in the low-luminosity regime. Pair production was calculated via direct integration of radiation spectra obtained from Monte Carlo radiative transfer in a post-processing manner. Their results are pair-free, consistent with our lowest accretion rate model. However, \citet{2011ApJ...735....9M} found that the pair production rate peaked in the midplane outside the horizon, while \citet{2021ApJ...907...73W} reported the highest pair production rates along the jet-disk boundary. In contrast, our model shows that the pair production rate depends strongly on the local proton scattering optical depth and thus peaks towards the midplane. This discrepancy could be attributed to the fact that pair production is fundamentally a global radiative process, while our approach approximates it through local processes. As mentioned in \citet{2011ApJ...735....9M}, Comptonized photons may be relativistically beamed, effectively enhancing pair production rates near the horizon.

\subsection{Insights from Kinetic Simulations} \label{subsec:insight}
%^^^^^^^^^^^^^^^^^^^^^^^^^^^^^^^^^^^^^^^^^^^^^^^^^^^^^^^^^^^^^^^^^^^^^^^^^^^^^^^^^^^^^^^^^^^^^^^^^^^^^^^^^^^^^^%
A number of recent kinetic studies have clarified some of the microphysical aspects governing pair processes in black hole coronae. \citet{2024NatCo..15.7026N} presented the first self-consistent simulation of a black hole corona in a local box with all relevant pair and radiative processes included. Aside from the externally supplied energy, all coronal properties, such as the scattering depth and magnetization, are determined self-consistently. Their results suggest that, in order to examine pair-regulated coronal physics in a global simulation, radiative cooling and two temperature physics are essential ingredients; \citet{2024PhRvL.132h5202G} showed that in strongly Alfvénic turbulence of pair plasma coupled to photons via Compton scattering, the pairs naturally develop an MeV-scale non-thermal tail. Although their calculation did not include pair processes or other radiative losses, the emergence of this high energy tail suggests that non-thermal processes may contribute to the pair production rate; \citet{2026arXiv260100518G} showed that in strongly turbulent electron–ion–photon plasma, a two‑temperature state develops self‑consistently, with electrons cooling rapidly via Compton losses to $\sim 100$\,keV $\sim 10^9$\,K. This result suggests that our adopted electron temperature of $T_e=10^9$\,K is a reasonable choice.

\subsection{Limitations} \label{subsec:limit}
%^^^^^^^^^^^^^^^^^^^^^^^^^^^^^^^^^^^^^^^^^^^^^^^^^^^^^^^^^^^^^^^^^^^^^^^^^^^^^^^^^^^^^^^^^^^^^^^^^^^^^^^^^^^^^^%

First, we consider only soft photon generation via thermal bremsstrahlung, but synchrotron emission could be an important source of soft photons \citep{1983AIPC..101..400T, 1983ApJ...269...49A, 1984ApJ...285..747B, 1999ApJ...517..381E}. In particular, \citet{1999ApJ...517..381E} showed that including synchrotron soft photons shifts the pair‑equilibrium curve toward smaller $\Theta_e$, thereby increasing the equilibrium $z$ value and enhancing the pair population within the domain. Additionally, non-thermal pair cascades may significantly contribute to producing pairs \citep{1985ApJ...294L..79Z, 1987MNRAS.227..403S, 1989ApJ...340..203T, 1995ApJ...440..100K}. Incorporating these two physical processes could drastically alter the pair physics timescale. Additionally, the drizzle model/kinetic process could be an important source of pair injection in hot, advective accretion flows and might contribute more significantly than advection. In particular, even in our current model where we set $n_p=0$ along the jet, so the pair-supply rate there is entirely due to advection --- which is less susceptible to uncertainties arising from numerical floors --- we still expect that kinetic processes will constitute an additional, and likely dominant, source of pairs for the bipolar jet. Our current implementation should be considered as a lower limit on the pair production rate.

Second, the radiative transfer model is not fully self-consistent. We adopt an ad hoc cooling function that instantaneously removes viscous dissipation and regulates the disk’s scale height. Also, the mass accretion rate has been prescribed independently, despite its expected dependence on $\dot{M}$. Additionally, pair production physics involves transferring energy from electrons to photons via Comptonization, enabling photons to reach energies sufficient for pair creation. These pairs subsequently annihilate, producing photons that can be absorbed or scattered by electrons. Such couplings are absent in our current model. The first‑order correction needed to incorporate such couplings is to evolve the individual species temperatures, adopt a lepton‑only cooling function, and include Coulomb energy exchange between leptons and hadrons. This can be achieved by separately evolving the entropy equations \citep{2015MNRAS.454.1848R, 2019MNRAS.486.2873C} for electrons, positrons, and ions.

Third, the pair production model is simplified. Specifically, the spectrum of pair-producing photons is assumed to comprise a flat continuum and a Wien component. This approximation should be reasonable in regions of higher optical depth and large Compton-$y$, where equilibrium between the electron gas and the photons interacting through Compton scattering can be established. However, it begins to break down in plasma with lower optical depth, where such equilibrium cannot be maintained. In these regions, non‑thermal processes may become important and produce photons that deviate from a thermal spectrum. Also note that this approximation would fail in regions of high absorption optical depth, i.e., near the midplane where the dense disk resides. However, by comparing the Coulomb‑collision and orbital time scales, we argue that in these regions the fluid remains single‑temperature and is sufficiently cold to suppress any pair production, and we disable pair production accordingly. Moreover, the Comptonization model is treated as a local approximation, whereas in practice, Compton scattering is inherently non-local at moderate scattering optical depths. Addressing these limitations would require coupling the pair production scheme with a radiative transfer model, which lies beyond the scope of this work.

%%%%%%%%%%%%%%%%%%%%%%%%%%%%%%%%%%%%%%%%%%%%%%%%%%%%%%%%%%%%%%%%%%%%%%%%%%%%%%%%%%%%%%%%%%%%%%%%%%%%%%%%%%%%%%%%%%%%%%%%%%%%%%%%%%%%%%%%%%%%%%%%%%%%%%
%Pair physics can be incorporated into full radiation transport \citep{1996ApJ...470..249P}, but doing so is beyond the scope of this paper. 
%%%%%%%%%%%%%%%%%%%%%%%%%%%%%%%%%%%%%%%%%%%%%%%%%%%%%%%%%%%%%%%%%%%%%%%%%%%%%%%%%%%%%%%%%%%%%%%%%%%%%%%%%%%%%%%%%%%%%%%%%%%%%%%%%%%%%%%%%%%%%%%%%%%%%%

%==============================================================================================================%

\section{Conclusion} \label{sec:conclu}

%%%%%%%%%%%%%%%%%%%%%%%%%%%%%%%%%%%%%%%%%%%%%%%%%%%%%%%%%%%%%%%%%%%%%%%%%%%%%%%%%%%%%%%%%%%%%%%%%%%%%%%%%%%%%%%%%%%%%%%%%%%%%%%%%%%%%%%%%%%%%%%%%%%%%%
%There has been increasing interest in the field to explore additional physics that regulate the thermal and dynamical properties of plasma in accretion flows around black holes.
%%%%%%%%%%%%%%%%%%%%%%%%%%%%%%%%%%%%%%%%%%%%%%%%%%%%%%%%%%%%%%%%%%%%%%%%%%%%%%%%%%%%%%%%%%%%%%%%%%%%%%%%%%%%%%%%%%%%%%%%%%%%%%%%%%%%%%%%%%%%%%%%%%%%%%

In this study, by performing a series of GRMHD simulations in the SANE state with pair physics incorporated, we investigate whether electron-positron pair production could be an additional physical process that regulates the thermal and dynamical properties of plasma in accretion flows around black holes. We model positrons as an additional passive variable that dynamically advects with the main accretion flow but maintains a fixed temperature. This approximation allows us to analyze the spatial distribution and relevant timescales in the simplest possible setting. 

The major results are that the pair-equilibrium timescale governs the spatial-temporal distribution of pairs - they remain in near-equilibrium only in the region that is moderately (scattering) optically thick ($\tau_{p} \sim \mathcal{O}(1)$). This region acts as a `pair source' that injects pairs into the surroundings, because the pair equilibrium timescale in such a region is short ($\sim \mathcal{O}(1)$\,$GM/c^{3}$); otherwise, the pair fraction is controlled purely by advection since the equilibrium timescale becomes long when $\tau_{p}$ is low. The strong advective flux generated by the radial infall of pairs within the ISCO can drive the local pair population out of equilibrium, even when $\tau_p$ is of order unity. Advection can inject pairs into the bipolar jets, thereby providing a pair density exceeding the Goldreich-Julian value along magnetically dominated jets. In regions of high $\tau_{p}$, pairs are predominantly produced through photon-photon collision processes. In contrast, pairs in regions of low $\tau_{p}$ are entirely produced by particle-particle interactions, which occur at a much slower rate than photon-photon processes. The mass accretion rate of the system influences the spatial location of the pair-maximized thin strip, thereby impacting the effectiveness of pairs being captured by outflows and, consequently, altering the efficiency of pair winds. Also, our results are generally not highly sensitive to variations in black hole mass, and we find that our choice of manually shutting down pair production based on $\tau_{\rm crit}$ is self-consistent.

%==============================================================================================================%

\section*{Acknowledgements}
Ho-Sang (Leon) Chan acknowledges support from the Croucher Scholarship for Doctoral Studies by the Croucher Foundation. MB and JD acknowledge support from NASA Astrophysics Theory Program grants 80NSSC22K0826 and 80NSSC24K1094 and Chandra award TM3-24003X.

\section*{Data Availability}

The simulation data underlying this article will be shared on reasonable request to the corresponding author.

%%%%%%%%%%%%%%%%%%%% REFERENCES %%%%%%%%%%%%%%%%%%

% The best way to enter references is to use BibTeX:

\bibliographystyle{mnras}
\bibliography{example} % if your bibtex file is called example.bib

% Alternatively you could enter them by hand, like this:
% This method is tedious and prone to error if you have lots of references
%\begin{thebibliography}{99}
%\bibitem[\protect\citeauthoryear{Author}{2012}]{Author2012}
%Author A.~N., 2013, Journal of Improbable Astronomy, 1, 1
%\bibitem[\protect\citeauthoryear{Others}{2013}]{Others2013}
%Others S., 2012, Journal of Interesting Stuff, 17, 198
%\end{thebibliography}

%%%%%%%%%%%%%%%%%%%%%%%%%%%%%%%%%%%%%%%%%%%%%%%%%%

\newpage 

%%%%%%%%%%%%%%%%% APPENDICES %%%%%%%%%%%%%%%%%%%%%

\appendix

\section{The Svensson Model} \label{app:svensson} 
%==============================================================================================================%
%#####################################################################################################################%
\begin{figure}
    \centering
    \includegraphics[width=1.0\linewidth]{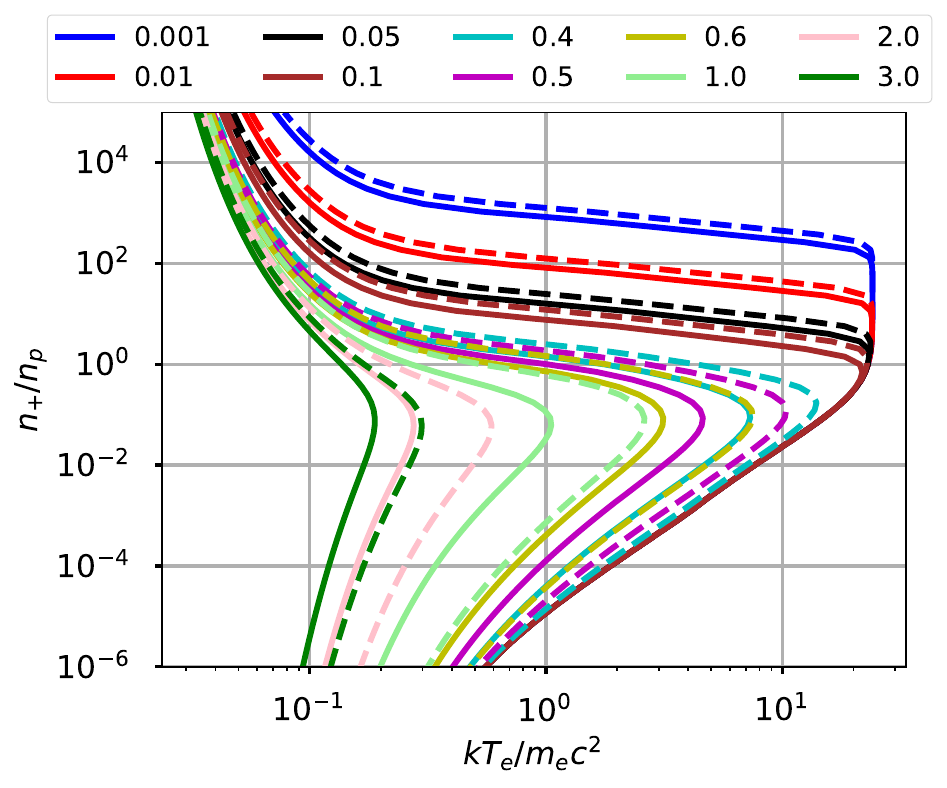}
    \caption{The one-zone equilibrium model presented in \citet{1984MNRAS.209..175S}, where we assume a plasma cloud with a size of $H = 3\times10^{7}$ cm. Here, lines of different colors represent various proton scattering optical depths, $\tau_{p}$. The dashed lines correspond to results with the geometric parameter $\xi = 1/3$, while the solid lines represent results with $\xi = 1$. \label{fig:svensson_equb}}
\end{figure}
%#####################################################################################################################%

We briefly review the one-zone equilibrium model presented in \citet{1984MNRAS.209..175S} and show the equilibrium $z - \Theta_{e}$ curve in Figure \ref{fig:svensson_equb}. In \citet{1984MNRAS.209..175S}, the results depend on a geometric parameter, $\xi$, which is set to $1/3$. For our study, we assume a slab geometry and thus set $\xi = 1$. We also include results with $\xi = 1/3$ in the same figure for reference, allowing comparison with previous work, including \citet{1991MNRAS.249..177B} and \citet{1999ApJ...517..381E}.

%==============================================================================================================%

\onecolumn

%==============================================================================================================%

\section{Pair Production Rate} \label{app:rates}

We briefly review the procedures for computing the pair production rates. We mostly follow \citet{1989ApJ...340.1024W} but also make references to \citet{1984MNRAS.209..175S}. First, the radiation field is assumed to be the sum of the flat, unscattered bremsstrahlung continuum $n_{F}(x)$, and a Wien peak $n_{W}(x)$
\begin{equation}
    n(x) = n_{F}(x) + n_{W}(x)
\end{equation}

Here, $n(x)$ is the number density of photons and the dimensionless frequency $x = h\nu/m_{e}c^{2}$ with $\nu$ being the photon frequency and $h$ the Planck constant. The bremsstrahlung continuum is given by
\begin{equation}
\begin{aligned}
    n_{F}(x) &= n_{1}\frac{1}{2}\text{exp}\left(-\frac{x}{\Theta_{e}}\right)\frac{1}{x}, \\
    n_{1} &= \frac{2}{\pi}\left(\frac{\alpha_{f}}{r_{e}}\right)^{3}x_{B}^{2}\Theta_{e}\left[\frac{1}{\text{ln}(\Theta_{e}/x_{B})} + \frac{y_{1}}{1+y_{1}}\right], 
\end{aligned}
\end{equation}
where $\alpha_{f}$ is the fine structure constant, $r_{e}$ is the classical electron radius, $x_{B}$ is the bremsstrahlung self-absorption dimensionless frequency, and $y_{1}$ is the Compton parameter. The Wien peak is given as
\begin{equation}
    n_{W}(x) = n_{\gamma}\frac{1}{2}\left(\frac{x}{\Theta_{e}}\right)^{2}\text{exp}\left(-\frac{x}{\Theta_{e}}\right)\frac{1}{\Theta_{e}}
\end{equation}
where the normalization constant $n_{\gamma}$ is given as
\begin{equation}
\begin{aligned}
    n_{\gamma}/t_{\text{esc}} &= f_{B}\dot{n}_{\gamma}^{B}, \\
    t_{\text{esc}} &= \frac{2H}{c}(1 + \tau_{W}(\Theta_{e})), \\
    \tau_{W} &= \tau_{T} 
    \begin{cases}
        1 + 5\Theta_{e} + 0.4\Theta_{e}^{2}, & \text{if } \Theta_{e} \leq 1, \\
        (3/16)\Theta_{e}^{-2}(\text{ln}2\eta\Theta_{e} + 3/4)(1 + 0.1/\Theta_{e})^{-1}, & \text{otherwise}.
    \end{cases}
\end{aligned}
\end{equation}

Here, $\dot{n}_{\gamma}^{B}$ is the total photon production rate by thermal bremsstrahlung. Also, $f_{B}$ is the fraction of photon produced via bremsstrahlung and compton up-scattered to the Wien peak. Additionally, $\eta = \text{exp}(-\gamma_{E})$ where $\gamma_{E}$ is the Euler constant. For pair productions, we consider photon-photon, photon-particle, and particle-particle processes. First, the pair production rate due to photon-photon collision both from the Wien peak is given as 
\begin{equation}
    (\dot{n}_{+})_{WW} = cr_{e}^{2}n_{\gamma}^{2} 
    \begin{cases}
        (1/8)\pi^{2}\Theta_{e}^{-3}\text{exp}(-2/\Theta_{e})(1 + 2.88\Theta_{e}^{0.934}), & \text{if } \Theta_{e} \leq 1, \\
        (1/2)\pi\Theta_{e}^{-2}\text{ln}(2\eta\Theta_{e} + 0.38), & \text{otherwise}.
    \end{cases}
\end{equation}

Also, the pair production rate due to collisions between Wien photons with those from the flat continuum is
\begin{equation}
    (\dot{n}_{+})_{WF} = cr_{e}^{2}n_{1}n_{\gamma}\frac{\pi^{2}}{4}\text{exp}\left(-\frac{2}{\Theta_{e}}\right)
\end{equation}

The photon-particle rate due to collisions between Wien photons and electrons/positrons is 
\begin{equation}
    (\dot{n}_{+})_{We} = \alpha_{f}cr_{e}^{2}n_{\gamma}(n_{+} + n_{e})
    \begin{cases}
        \frac{4\pi}{27}\text{exp}\left(-\frac{2}{\Theta_{e}}\right)(1 + 5\Theta_{e} + 0.4\Theta_{e}^{2}), & \text{if } \Theta_{e} \leq 0.18, \\
        \frac{4\pi}{27}\text{exp}\left(-\frac{2}{\Theta_{e}}\right)[(3/16)\Theta_{e}^{-2}(\text{ln}2\eta\Theta_{e} + 3/4)(1 + 0.1/\Theta_{e})^{-1}], & \text{if } 0.18 \leq \Theta_{e} \leq 2, \\
        \left(\frac{56}{9}\text{ln}(2\eta\Theta_{e}) - \frac{8}{27}\right)(1 + 0.5/\Theta_{e})^{-1},& \text{otherwise}.
    \end{cases}
\end{equation}

The photon-particle rate due to collisions between Wien photons and protons is 
\begin{equation}
    (\dot{n}_{+})_{Wp} = \alpha_{f}cr_{e}^{2}n_{\gamma}n_{p}
    \begin{cases} 
        \pi\Theta_{e}\text{exp}(-2/\Theta_{e})(1 + 0.9\Theta_{e})^{-1}, & \text{if } \Theta_{e} \leq 2, \\
        \frac{28}{9}\text{ln}(2\eta\Theta_{e} + 1.7) - \frac{92}{27}, & \text{otherwise}.
    \end{cases}
\end{equation}

We consider only the electron-electron collision rate since the proton-electron rate is always lower \citep{1982ApJ...258..335S}. The electron-electron rate is given as
\begin{equation}
    (\dot{n}_{+})_{ee} = cr_{e}^{2}(n_{e} + n_{+})^{2}
    \begin{cases} 
        \frac{112}{27\pi}\alpha_{f}^{2}(\text{ln}\Theta_{e})^{3}(1 + 0.058/\Theta_{e})^{-1}, & \text{if } \Theta_{e} > 3.978 , \\
        2 \times 10^{-4}\Theta_{e}^{3/2}\text{exp}(-2\Theta_{e})(1 + 0.015\Theta_{e}), & \text{if } \Theta_{e} < 0.383 .
    \end{cases}
\end{equation}
note that to our knowledge there are no analytic approximation in the range of $\Theta_{e} \sim 1$. We therefore perform an analytic fit to bridge between the above formulae. The fit is given by a $5$th-order polynomial:
\begin{equation}
\begin{aligned}
    x_{t} &= \rm{log}_{10}(\Theta_{e}), \\
    y_{t} &= \sum_{n = 0}^{5}a_{n}x_{t}^{n}, \\
    (\dot{n}_{+})_{ee} &= cr_{e}^{2}(n_{e} + n_{+})^{2}10^{y_{t}}
\end{aligned}
\end{equation}
in the range of $0.383 \leq \Theta_{e} \leq 3.978$. Here, $a_{5} = 0.262$, $a_{4} = -1.262$, $a_{3} = 2.137$, $a_{2} = -1.949$, $a_{1} = 2.655$, and $a_{0} = -4.948$. The total pair production rate $\dot{n}_{C}$ and the pair annihilation rate $\dot{n}_{A}$ are given as 
\begin{equation}
\begin{aligned}
    \dot{n}_{C} = (\dot{n}_{+})_{WW} + (\dot{n}_{+})_{WF} + (\dot{n}_{+})_{We} + (\dot{n}_{+})_{Wp} + (\dot{n}_{+})_{ee}, \\
    \dot{n}_{A} = \pi cr_{e}^{2}n_{e}n_{+}\left\{1 + \left[\frac{2\Theta_{e}^{2}}{\text{ln}(1.12\Theta_{e} + 1.3)}\right]\right\}^{-1}.
\end{aligned}
\end{equation}

%==============================================================================================================%

\section{Radiative Processes} \label{app:radiative}

We briefly review the radiative processes involved in pair production. We mostly follow \citet{1989ApJ...340.1024W} but also make reference to \citet{1984MNRAS.209..175S, 1996ApJ...465..312E, 1999ApJ...517..381E}.

%==============================================================================================================%

\subsection{Bremsstrahlung} \label{subapp:brem}

The thermal bremsstrahlung emissivity contains contributions from electron/positron-proton $\dot{n}_{ep}(x)$, electron-electron/positron-positron $\dot{n}_{ee}(x)$, and electron-positron $\dot{n}_{+-}(x)$ interactions, which are given as
\begin{equation}
\begin{aligned}
    \dot{n}_{ep}(x) &= (1 + 2z)\text{ln}\left[4\eta(1+3.42\Theta_{e})\frac{\Theta_{e}}{x}\right](1 + 2\Theta_{e} + 2\Theta_{e}^{2})\dot{n}_{0}(x), \\
    \dot{n}_{ee}(x) &= [z^{2} + (1+z)^{2}]\text{ln}\left[4\eta(11.2 + 10.4\Theta_{e}^{2})\frac{\Theta_{e}}{x}\right](\frac{3}{5}\sqrt{2}\Theta_{e} + 2\Theta_{e}^{2})\dot{n}_{0}(x), \\
    \dot{n}_{+-}(x) &= z(1 + z)\text{ln}\left[4\eta(1+10.4\Theta_{e}^{2})\frac{\Theta_{e}}{x}\right]2(\sqrt{2} + 2\Theta_{e} + 2\Theta_{e}^{2})\dot{n}_{0}(x), \\
    \dot{n}_{0}(x) &= \frac{16}{3}\alpha_{f}cr_{e}^{2}n_{p}^{2}\left[\text{exp}\left(\frac{1}{\Theta_{e}}\right)K_{2}\left(\frac{1}{\Theta_{e}}\right)\right]^{-1}x^{-1}\text{exp}\left(-\frac{x}{\Theta_{e}}\right)
\end{aligned}
\end{equation}
where $K_{2}$ is the modified Bessel function of the second kind of order $2$. The total bremsstrahlung production rate is obtained by integrating the spectrum beyond $x_{B}$ and is given as
\begin{equation}
\begin{aligned}
    \dot{n}_{\gamma}^{B} &= \frac{16}{3}\alpha_{f}cr_{e}^{2}n_{p}^{2}\left[\text{exp}\left(\frac{1}{\Theta_{e}}\right)K_{2}\left(\frac{1}{\Theta_{e}}\right)\right]^{-1}\text{ln}\left(\frac{\Theta_{e}}{x_{B}}\right)S(z,\Theta_{e}), \\
    S(z,\Theta_{e}) &= (1 + 2z)\text{ln}\left[4\eta(1+3.42\Theta_{e})\left(\frac{\Theta_{e}}{x_{B}}\right)^{1/2}\right](1 + 2\Theta_{e} + 2\Theta_{e}^{2}) \\
    &+ [z^{2} + (1+z)^{2}]\text{ln}\left[4\eta(11.2 + 10.4\Theta_{e}^{2})\left(\frac{\Theta_{e}}{x_{B}}\right)^{1/2}\right](\frac{3}{5}\sqrt{2}\Theta_{e} + 2\Theta_{e}^{2}) \\
    &+ z(1 + z)\text{ln}\left[4\eta(1+10.4\Theta_{e}^{2})\left(\frac{\Theta_{e}}{x_{B}}\right)^{1/2}\right]2(\sqrt{2} + 2\Theta_{e} + 2\Theta_{e}^{2}).
\end{aligned}
\end{equation}

The bremsstrahlung self-absorption dimensionless frequency $x_{B}$ is obtained by solving the non-linear equation
\begin{equation}
\begin{aligned}
    (n_{+} + n_{e})\sigma_{T} &= a_{\rm abs}(x_{B})\frac{\tau_{T}(1 + \tau_{T})}{1 + \tau_{T}^{2}\text{min}[1,8\Theta_{e}]}, \\
    a_{\rm abs}(x) &= \frac{1}{c}\frac{\dot{n}_{ep}(x) + \dot{n}_{ee}(x) + \dot{n}_{+-}(x)}{n_{BB}(x)}, \\
    n_{BB}(x) &= \lambda_{c}^{-3}\pi^{-2}x^{2}[\text{exp}(x/\Theta_{e})-1]^{-1}.
\end{aligned}
\end{equation}
where $\lambda_{c}$ is the reduced Compton wavelength.

%==============================================================================================================%

\subsection{Comptonization} \label{subapp:compton}

The normalized probability distribution for a photon to undergo $N$ scatterings before escaping is given as
\begin{equation}
\begin{aligned}
    P(N) &= \frac{1}{\tau_{T}^{2}}\text{exp}(-\frac{N}{\tau_{T}^{2}}), \\
    N &= \frac{\text{ln}(x_{f}/x_{i})}{\text{ln}(1 + 4\Theta_{e} + 16\Theta_{e}^{2})}
\end{aligned}
\end{equation}
where $x_{i}$ ($x_{f}$) is the initial (final) energy of the photon. The fraction of photons with an initial energy $x$ that are Compton up-scattered to a final energy greater than $\Theta_{e}$ before escaping is thus
\begin{equation}
    f(x) = \int_{N}^{\infty}P(y)d(y) = \text{exp}\left[-\frac{\text{ln}(\Theta_{e}/x)}{\tau_{T}^{2}(1 + 4\Theta_{e} + 16\Theta_{e}^{2})}\right]
\end{equation}

The fraction of bremsstrahlung photons that are up-scattered is thus 
\begin{equation}
\begin{aligned}
    f_{B} &= \int_{x_{B}}^{\Theta_{e}}f(x)2\frac{\text{ln}(\Theta_{e}/x)}{\text{ln}^{2}(\Theta_{e}/x_{B})}\frac{dx}{x} = 2[y_{1}^{2} - (y_{1} + y_{1}^{2})e^{-1/y_{1}}], \\
    y_{1} &= \tau_{T}^{2}\frac{\text{ln}(1 + 4\Theta_{e} + 16\Theta_{e}^{2})}{\text{ln}(\Theta_{e}/x_{B})}
\end{aligned}
\end{equation}

%==============================================================================================================%

\subsection{Weak Dependence on Black Hole Mass} \label{subapp:dependence}

We argued in Section \ref{subsec:bhmass} that our results generally depend only weakly on the black hole mass. To demonstrate this, we extend the argument presented in \citet{1986AcA....36..255S}. In the thin strip where the pair density is maximized, and at fixed temperature, the pair equilibrium equation can be written as $n_{\gamma}^{2} = n_{p}(2z+1)\Xi(\Theta_{e})$, where $\Xi$ is a function that depends on $\Theta_{e}$ only. We adopt the approximation $(\dot{n}_{+})_{WW} \approx \dot{n}_{A}$, assuming that pair production is dominated by photon-photon collisions. Next, we note that $n_{\gamma} = f_{B}[z, \Theta_{e}, \ln(\Theta_{e}/x_{B})]\,\dot{n}_{\gamma}^{B}\,(2H/c)\,[1 + \tau_{T}\varphi(\Theta_{e})]$,
where the fraction of upscattered photons $f_{B}$ depends on $z$, $\Theta_{e}$, and $\ln(\Theta_{e}/x_{B})$, and $\varphi(\Theta_{e})$ is a function of $\Theta_{e}$ alone.
We first show that $\tau_{T}$ is independent of the black hole mass. Recall that $\tau_{T} = 2\sigma_{T}Hn_{p}(2z+1)$, where $H \sim L_{\rm Unit}$ and $n_{p} \sim M_{\rm Unit}/L_{\rm Unit}^{3}$. Since $M_{\rm Unit}$ is defined such that $\dot{M}_{\rm EH} M_{\rm Unit}/T_{\rm Unit} \propto \frac{4\pi G M_{\rm BH}m_{p}c}{\sigma_{T}}$, where $\dot{M}_{\rm EH}$ is the mass accretion rate measured at the event horizon, it follows that $M_{\rm Unit}/L_{\rm Unit}^{2}$ is constant. Therefore, $\tau_{T}$ is independent of black hole mass. We also note that $\dot{n}_{\gamma}^{B} = n_{p}^{2}\Psi[z, \Theta_{e}, \ln(\Theta_{e}/x_{B})]$, where $\Psi$ is a function of $z$, $\Theta_{e}$, and $\ln(\Theta_{e}/x_{B})$. Substituting, we obtain $f_{B}[z, \Theta_{e}, \ln(\Theta_{e}/x_{B})]\,(2/c)\,[1 + \tau_{T}\varphi(\Theta_{e})]\,n_{p}H = \frac{2z+1}{\Psi[z, \Theta_{e}, \ln(\Theta_{e}/x_{B})]}$. Since $n_{p}H \sim M_{\rm Unit}/L_{\rm Unit}^{2}$ is constant, the pair equilibrium equation can be rearranged as
$\frac{(2z + 1)}{\Psi[z, \Theta_{e}, \ln(\Theta_{e}/x_{B})]\,f_{B}[z, \Theta_{e}, \ln(\Theta_{e}/x_{B})]\,[1 + \tau_{T}\varphi(\Theta_{e})]} = \frac{2}{c}n_{p}H$,
where the right-hand side is independent of black hole mass. The only mass dependence enters through the self-absorption frequency $x_{B}$, which scales approximately as $1/\sqrt{H}$ \citep{1986AcA....36..255S}. Consequently, $\ln(\Theta_{e}/x_{B})$ depends only weakly on $H$, explaining the weak dependence of $z$ on black hole mass, which sets the value of $H$ via $L_{\rm Unit}$. In the upper corona, local pair processes are negligible and are instead governed entirely by advection, which is controlled by the dynamics and is scale-free.

%%%%%%%%%%%%%%%%%%%%%%%%%%%%%%%%%%%%%%%%%%%%%%%%%%

% Don't change these lines
\bsp	% typesetting comment
\label{lastpage}
\end{document}